\documentclass[pra,showpacs,aps,amsmath,amssymb]{revtex4-1}
\usepackage{graphicx}
\usepackage{graphics}
\usepackage{amsmath,amsfonts,amssymb}
\usepackage{bm}
\usepackage{epsfig}
\usepackage[english,french]{babel}
\usepackage[utf8]{inputenc}
\usepackage{xcolor}
\usepackage{ulem}

\begin{document}

\title{Reformulation of the strong field approximation for light-matter interactions}

\author{A. Galstyan$^1$\footnote{alexander.galstyan@uclouvain.be}, O. Chuluunbaatar$^{2,3}$,
              A. Hamido$^1$, Yu. V. Popov$^{4,2}$,\\ F. Mota-Furtado$^5$, P. F. O'Mahony$^5$,  N. Janssens$^1$,
              F. Catoire$^6$ and B. Piraux$^1$}

\affiliation{
$^1$Institute of Condensed Matter and Nanosciences, Universit\'e Catholique de Louvain,\\ 
         2 chemin du cyclotron, Box L7.01.07,  B-1348 Louvain-la-Neuve, Belgium\\
$^2$Joint Institute for Nuclear Research, Dubna, Russia\\
$^3$School of Mathematics and Computer Science, National University of Mongolia, UlaanBaatar, Mongolia\\
$^4$Skobeltsyn Institute of Nuclear Physics, Moscow State University, Moscow, Russia\\
$^5$Department of Maths, Royal Holloway, University of London, Egham, Surrey   TW20 0EX, United Kingdom\\
$^6$Centre des Lasers Intenses et Applications, UMR 5107, CNRS-CEA-Universit\'e de Bordeaux, \\
         351 Cours de la Lib{\'e}ration, Talence F-33405, France }

\begin{abstract}
We consider the interaction of hydrogen-like atoms with a strong laser field and show that the strong field approximation and all its
variants may be grouped into a set of families of approximation schemes. This is done by introducing an ansatz describing
the electron wave packet as the sum of the initial state wave function times a phase factor and a function which is the perturbative
solution in the Coulomb potential of an inhomogeneous time-dependent Schr\"odinger equation. It is the phase factor that characterizes
a given family. In each of these families,  the velocity and length gauge version of the approximation scheme lead to the same results
at each order in the Coulomb potential. By contrast, irrespective of the gauge, approximation schemes belonging to different families
give different results. Furthermore, this new  formulation of the strong field approximations allows us to gain deeper insight into the 
validity of the strong field approximation schemes. In particular, we address two important questions:  the role of the Coulomb potential 
in the output channel and the convergence of the perturbative series in the Coulomb potential. In all the physical situations we consider
here, our results are compared to those obtained by solving numerically the time-dependent Schr\"odinger equation.
\end{abstract}


\maketitle

\section{INTRODUCTION}

The solution of the time-dependent Schr\"odinger equation (TDSE) that describes the interaction of an intense electromagnetic pulse 
with an atom or a molecule is a very challenging problem from the numerical point of view. This is particularly true in the long wavelength 
limit where the numerical calculations become rapidly intractable even for one-active electron systems. In this situation, one has to rely
on analytical or semi-analytical models that, in addition, offer the advantage of providing valuable insight into the physical mechanisms 
that underly the interaction processes.\\

The first model that laid the foundation of our understanding of laser-atom interactions has been developed by Keldysh. In 1965, he 
published a seminal paper \cite{Keldysh65} on the nonlinear ionization of atoms and dielectrics by  strong  electromagnetic waves. 
This model describes, in the length gauge, the transition of an atom, from an initially unperturbed bound state to a dressed continuum state. 
This dressed continuum is a Volkov state \cite{volkov35}. It takes into account the electron-field interaction at all orders while neglecting completely 
the Coulomb potential. Keldysh introduced the adiabaticity parameter $\gamma=\omega\sqrt{2I_p}/E$ where, in atomic units, $\omega$ 
is the photon energy, $I_p$ the ionization potential and $E$ the electric field amplitude. He showed that tunneling and multiphoton ionization 
are actually two limiting cases of nonlinear ionization: tunneling is the dominant process for $\gamma\ll 1$ while for $\gamma\gg 1$, ionization
proceeds by the absorption of many photons.\\

Soon after Keldysh's contribution, Perelomov, Popov and Terent'ev (PPT) published a series of papers \cite{PPT1,PPT2,PPT3}. Their model is
based on the approximate solution, in the length gauge, of a time-dependent Lippmann-Schwinger like equation. It involves a Green's function 
which is  expressed in terms of the classical action and it is assumed that the initial state is hardly depleted. For electrons bound by short range 
forces, the PPT approach reproduces the Keldysh theory. In the limit where the Coulomb field is much smaller than the external field so that it can 
be regarded as a small perturbation, effects due to the long range of the Coulomb potential are accounted for in the classical action \cite{PPT3}. In 
addition, and contrary to Keldysh theory, the ionization rate, averaged over the laser period, tends to the correct value in the adiabatic (quasi-static)
limit.\\

Later, when the first experimental data on the above-threshold ionization spectra became available, Reiss \cite{Reiss80} and Faisal \cite{Faisal73} 
developed an approach based on the S-matrix theory in the velocity gauge. It is after these papers that the terminology ``strong field approximation'' 
(SFA) appeared. In fact, what is understood as the SFA, is the first term of the perturbative series in the Coulomb potential associated to the S-matrix. 
High order terms of SFA describe multiple re-scattering of the electron by the ionic core \cite{lewenstein_rings} as this electron undergoes a quiver motion 
driven by the external electric field. It is important to note that the S-matrix treatment can also be developed in the length gauge by using the 
$\vec{E}\cdot\vec{r}$ form of the interaction Hamiltonian. In that case one obtains the Keldysh result at the first order.\\

SFA has been widely used to describe the mechanism for many strong field phenomena such as high order harmonic generation \cite{Lewenstein1}, 
high energy  electron spectra and even processes involving more than one electron like non-sequential double photoionization \cite{krausz09}. 
However, SFA has a serious flaw: the length and velocity gauge versions of SFA give results that may differ significantly \cite{bergues07}. This problem 
is still unsolved. Within this context, it is interesting to mention here three commonly accepted viewpoints which are nicely summarized in 
\cite{popr}: (i) the problem of the gauge invariance is misunderstood and a proper treatment of the full  ``atom+field''  Hamiltonian should make 
the velocity and the length gauge versions of the SFA amplitude equivalent; (ii) there is a preferable gauge (depending on the physical process under 
consideration) which should be used in the calculations;  (iii) in the present stage, the theory is essentially non-invariant and should be modified to 
restore the gauge invariance property.\\
 
In attempting to modify the SFA theory along the lines of the third viewpoint, Faisal \cite{faisal07} managed to reformulate his S-matrix based approach and 
obtained a new gauge invariant SFA. In fact, his new SFA is identical to the old length-gauge SFA but differs from the old velocity-gauge SFA. This might also
indicate that the length gauge is the preferable one. However, we show that it is not the case. \\

In this contribution, we consider the interaction of atomic 
hydrogen with a strong pulsed oscillating field and we develop a rather  transparent approach that shows that the different SFA schemes can be grouped into 
a set of different families. In a given family, the length and the velocity SFA schemes give the same results at all orders in the Coulomb potential. However, 
irrespective of the gauge, two SFA schemes belonging to different families give different results. The two gauge invariant SFA schemes obtained 
by Faisal belong to the same family. The main idea of the present method is to introduce an ansatz that describes the electron wave packet as the sum of the 
initial wave function times a phase factor and a function which is the solution of an inhomogeneous TDSE that can be solved iteratively to generate a perturbative 
series in the Coulomb potential. It is essentially the phase factor that defines a given family. Furthermore, this simple reformulation of the SFA schemes allows 
us to address the two following important questions regarding the validity of the SFA schemes. The first one concerns the final state on which the SFA wave 
packet should be projected on to get the electron energy spectrum. In principle, this final state must be a Coulomb wave. However, in order to avoid enormous 
difficulties from the analytical point of view, Coulomb waves are usually replaced by plane waves. This further approximation is questionable in the case 
of very low energy electrons. The second important  question concerns the convergence of the perturbative series in the Coulomb potential. We address this 
question by solving numerically and iteratively the inhomogeneous TDSE and show in one particular case how the electron energy spectrum changes after 
the inclusion of increasing order terms.  In all cases, the SFA results are compared to those obtained by solving numerically the TDSE.\\

This paper is organized as follows.  Section II is devoted to the theory. We first transform the TDSE into a time-dependent 
Lippmann-Schwinger like equation. It is this equation that is the starting point of the PPT approach. The problems inherent to that equation are discussed in 
detail and justify the introduction of the particular ansatz we use to generate various families of SFA schemes. We focus on two particular families and 
establish their connection to  existing SFA schemes. The third section is devoted to the numerical treatment of the TDSE 
within the SFA schemes. Finally, before concluding, we compare two schemes based on an ansatz belonging to different families. This is done by calculating 
in a realistic situation the electron energy distribution along the polarization axis and by comparing the SFA results to those obtained by solving the TDSE.\\

Atomic units ($\hbar=e=m_e=1$) combined with the gaussian system for the electromagnetic field are used throughout unless otherwise 
specified.

\section{THEORY}

\subsection{Useful definitions}

In this contribution, we consider  fields linearly polarized along the z-axis and assume that the dipole approximation is valid. At this stage, we define the
vector potential and the electric field for a given time $t$ as follows:
\begin{eqnarray}
\label{eq_field}
\frac{1}{c}\vec{A}(t)&=&-b'(t)\vec{e},\\ 
\vec{\mathcal{E}}(t)&=&b''(t)\vec{e},
\end{eqnarray}
where $b(t)$, the expression of which will be specified later, vanishes  for $t\leq 0$ and $t\geq T$, with $T$ being the total pulse duration. $\vec{e}$ is a unit 
vector along the z-axis. It is also convenient to define the following quantity:
\begin{equation}
\label{eq_field_zeta}
\zeta(t)=\frac{1}{2}\int_0^t \mathrm{d}\xi [b'(\xi)]^2,
\end{equation}
and to write the canonical momentum $\vec{P}(t)$ and the action $S(\vec{p},t)$ in terms of $b(t)$:
\begin{eqnarray}
\label{eq_canon}
\vec{P}(t)&=&\vec{p}-b'(t)\vec{e},\\
S(\vec{p},t)&=&\frac{1}{2}\int_0^t  P^2(\xi)\mathrm{d}\xi.
\end{eqnarray}
where $\vec{p}$ is the electron momentum. In the following, we use a tilde accent for quantities defined in the configuration space and no accent if the 
same quantity is defined in momentum space.

\subsection{Time-dependent perturbative treatment of the Schr\"odinger equation }

The TDSE describing the interaction of a pulsed electric field with an hydrogenic system of nuclear charge Z and initially in its ground state $\tilde\varphi_0(\vec{r})$, is 
in the velocity (V) gauge, 
\begin{equation}
\label{eq_tdse_vg}
\left[\mathrm{i}\frac{\partial}{\partial t}+\frac{1}{2}\triangle_r+\frac{Z}{r}-\mathrm{i}b'(t)(\vec{e}\cdot\vec{\nabla}_r)-\zeta'(t)\right]\tilde\Phi_V(\vec{r},t)=0,
\;\;\;\;\;\; \tilde\Phi_V(\vec{r},0)=\tilde\varphi_0(\vec{r}),
\end{equation}
and in the length (L) gauge,
\begin{equation}
\label{eq_tdse_lg}
\left[\mathrm{i}\frac{\partial}{\partial t}+\frac{1}{2}\triangle_r+\frac{Z}{r}-b''(t)(\vec{e}\cdot\vec{r})\right]\tilde\Phi_L(\vec{r},t)=0,
\;\;\;\;\;\; \tilde\Phi_L(\vec{r},0)=\tilde\varphi_0(\vec{r}).
\end{equation}
It is important to note that in the V-gauge version of the TDSE, the interaction Hamiltonian includes the term $\zeta'(t)$ that is proportional to the square of 
the vector potential. In principle, and within the dipole approximation, this term can be eliminated by a simple unitary transformation of the solution $\tilde\Phi_V(\vec{r},t)$. 
However, this is only true for the exact solution. In the case of the SFA for instance, neglecting this term may lead to wrong results \cite{Reiss89}.
Because of the gauge invariance, these two TDSE are equivalent. Their solutions are related by the well known G\"oppert-Mayer gauge transformation:
\begin{equation}
\label{eq_GM_transf}
\tilde\Phi_L(\vec{r},t)=e^{-\mathrm{i}b'(t)(\vec{e}\cdot\vec{r})}\tilde\Phi_V(\vec{r},t).
\end{equation}
In the case of weak fields, we could treat the dipole interaction Hamiltonian as a perturbation. Instead, we assume that the electric field is very strong
compared to the Coulomb field and consider the Coulomb potential as a perturbation. In the following, we first consider the V-gauge and rewrite Eq. (\ref{eq_tdse_vg}) as 
follows:
\begin{equation}
\label{eq_tdse_vg1}
\left[\mathrm{i}\frac{\partial}{\partial t}+\frac{1}{2}\triangle_r- \mathrm{i}b'(t)(\vec{e}\cdot\vec{\nabla}_r)-\zeta'(t)\right] \tilde\Phi_{V}(\vec{r},t)=-\frac{Z}{r}\tilde\Phi_{V}(\vec{r},t). 
\end{equation}
If the right hand side of this equation is set equal to zero, the solution of the remaining equation is a Volkov wave function given by:
\begin{equation}
\label{eq_volkov_vg}
\tilde\chi_V(\vec{r},\vec{p},t)= e^{[\mathrm{i}\vec{p}\cdot\vec{r}-\mathrm{i}p^2t/2+\mathrm{i}b(t)(\vec{e}\cdot\vec{p})-\mathrm{i}\zeta(t)]}=e^{\mathrm{i}\vec{p}
\cdot\vec{r}-\mathrm{i}S(\vec{p},t)}.
\end{equation}
In terms of these Volkov wave functions, the  corresponding time-dependent Green's function is:
\begin{equation}
\label{eq_green_vg}
\tilde G_V(\vec{r},t; \vec{r}\;',t')=-\mathrm{i}\theta(t-t')\int\frac{\mathrm{d}^3p}{(2\pi)^3}\tilde\chi_V(\vec{r},\vec{p},t)\tilde\chi_V^*(\vec{r}\;',\vec{p},t').
\end{equation}
The general solution of Eq. (\ref{eq_tdse_vg1}) may be written as follows:
\begin{eqnarray}
\label{eq_phi_vg_green}
\tilde\Phi_V(\vec{r},t)&=&\mathrm{i}\int \mathrm{d}^3 r' \tilde G_V(\vec{r},t;\vec{r}\;',0)\tilde\varphi_0(\vec{r}\; ') -Z \int_0^t \mathrm{d}t'\ \int
\frac{\mathrm{d}^3 r'}{r'} \tilde G_V(\vec{r},t; \vec{r}\;',t')\tilde\Phi_V(\vec{r}\;',t');\\
&=&I_V^{(0)}+I_V.
\end{eqnarray}
Substituting $\tilde G_V(\vec{r},t;\vec{r}\;',0)$ in (\ref{eq_phi_vg_green}) by the expression (\ref{eq_green_vg}), we obtain:
\begin{eqnarray}
\label{eq_Iv0}
I_V^{(0)}&=&\theta(t)\int\frac{\mathrm{d}^3p}{(2\pi)^3}\tilde\chi_V(\vec{r},\vec{p},t)\varphi(\vec{p});\\
\label{eq_Iv}
I_V&=&iZ\int\frac{\mathrm{d}^3p}{(2\pi)^3}\tilde\chi_V(\vec{r},\vec{p},t)\int_0^t \mathrm{d}t'\int \frac{\mathrm{d}^3 r'}{r'}\tilde\chi^*_V(\vec{r}\;',\vec{p},t')\tilde\Phi_{V}(\vec{r}\ ',t').
\end{eqnarray}
Eq. (\ref{eq_phi_vg_green}) is an integral equation of Lippmann-Schwinger type. It can be solved by iteration, starting with $I_V^{(0)}$ as the zeroth order term to generate a perturbative
(Born) series in the Coulomb potential. It is important to note, at this stage, that each term, denoted by $I_V^{(n)}$, of this perturbative series is gauge invariant. In 
order to show it, we could start from Eq. (\ref{eq_tdse_lg}) which is the L-gauge version of Eq. (\ref{eq_tdse_vg}) and proceed as we did above. Instead, we can replace the Volkov wave function 
$\tilde\chi_V(\vec{r},\vec{p},t)$ in Eq. (\ref{eq_green_vg})-(\ref{eq_Iv}) by its L-gauge version $\tilde\chi_L(\vec{r},\vec{p},t)$. Since the V-gauge and L-gauge expressions of a Volkov wave 
function can be obtained one from the other by the usual G\"oppert-Mayer gauge transformation :
\begin{equation}
\label{eq_volkov_lg}
\tilde\chi_L(\vec{r},\vec{p},t)= e^{-\mathrm{i}b'(t)(\vec{e}\cdot\vec{r})}\tilde\chi_V(\vec{r},\vec{p},t)=e^{\mathrm{i}\vec{P}(t)\cdot\vec{r}-\mathrm{i}S(\vec{p},t)},
\end{equation}
and that:
\begin{equation}
\tilde\chi^*_L(\vec{r}\;',\vec{p},t')\tilde\chi_L(\vec{r}\;',\vec{p}\;',t')=\tilde\chi^*_V(\vec{r}\;',\vec{p},t')\tilde\chi_V(\vec{r}\;',\vec{p}\;',t'),
\end{equation}
one can easily see from Eq. (\ref{eq_Iv}) that any Born term $I^{(n)}_L$ in the L-gauge is related to corresponding term $I_V^{(n)}$ in the V-gauge by the usual gauge 
transformation: 
\begin{equation}
I_L^{(n)}=e^{-\mathrm{i}b'(t)(\vec{e}\cdot\vec{r})}I_V^{(n)}
\end{equation}
Let us now discuss in more detail the pertinence of the integral equation (\ref{eq_phi_vg_green}) and its perturbative (Born) series. A closer look to the first term $I_V^{(0)}$ in the
integral equation (\ref{eq_phi_vg_green}) shows that it undergoes a rapid dispersion due to the presence of the $e^{-\mathrm{i}p^2t/2}$ term in the expression (\ref{eq_volkov_vg}) of the Volkov 
wave function $\tilde\chi_V(\vec{r},\vec{p},t)$. In addition, since  this exponential term $e^{-\mathrm{i}p^2t/2}$ does not depend on the external field, it means that 
this rapid dispersion occurs even in the absence of the external field. In fact, in the absence of the field, we can rewrite the integral equation (\ref{eq_phi_vg_green}) in momentum
space as follows:
\begin{equation}
\label{eq_phi_vpt}
\Phi_V(\vec{p},t)=e^{-\mathrm{i}p^2t/2}\left[\varphi_0(\vec{p})+4\pi\mathrm{i}Z\int_0^t \mathrm{d}\xi\ e^{\mathrm{i}p^2\xi/2}\int\frac{\mathrm{d}^3p'}{(2\pi)^3}
\frac{\Phi_V(\vec{p}\;',\xi)}{|\vec{p}-\vec{p}\;'|^2}\right].
\end{equation}
If we solve iteratively this equation, we clearly see that each term of the perturbation expansion of $\Phi_V(\vec{p},t)$ is affected by a rapid dispersion. This may
be justified when $\varphi_0(\vec p)$ is an arbitrary wave packet. In our case however, the atom is and must stay in a stationary state in  the absence of an external field. 
One can prove that $\Phi_V(\vec p,t) = \varphi_0(\vec p)e^{-\mathrm{i}\varepsilon_0t}$ satisfies  Eq. (\ref{eq_phi_vpt}) which cannot be solved iteratively.\\

There are at least two ways of circumventing this difficulty. The first one is to follow the PPT approach. The starting point of this approach is the L-gauge version 
of the integral equation (\ref{eq_phi_vg_green}). Since the first term is affected by a rapid dispersion, it does not contribute to the current and can be neglected. Furthermore, the exact 
wave packet  that enters in the second term is replaced by the initial state wave function under the assumption that the mean time of ionization is much larger 
than the atomic times. The second way consists in introducing an ansatz to describe the exact wave packet as the sum of the initial stationary state times a phase 
factor and a function that is the solution of an inhomogeneous TDSE. \\

\subsection{Ansatz-based approach}

Let us first consider the V-gauge. The ansatz we introduce consists in separating the initial bound state component from the other components of the electron wave 
packet $\tilde\Phi_V(\vec{r},t)$. We write:
\begin{equation}
\label{eq_phi_V1}
\tilde\Phi_V(\vec r,t)=e^{-\mathrm{i}\varepsilon_0 t}\tilde\varphi_0(\vec{r})+\tilde F_V(\vec{r},t),
\end{equation}
where $\varepsilon_0=-Z^2/2$ is the ground state energy. After substitution in Eq. (\ref{eq_tdse_vg}), we obtain the following inhomogeneous TDSE for $\tilde F_V(\vec{r},t)$:
\begin{equation}
\label{eq_inh_tdse_vg}
\left[\mathrm{i}\frac{\partial}{\partial t}+\frac{1}{2}\triangle_r- \mathrm{i}b'(t)(\vec{e}\cdot\vec{\nabla}_r)-\zeta'(t)\right] \tilde F_{V}(\vec{r},t)=-\frac{Z}{r}\tilde F_{V}(\vec{r},t)+\left[\mathrm{i}b'(t)(\vec{e}\cdot\vec{\nabla}_r)+\zeta'(t)\right]e^{-\mathrm{i}\varepsilon_0 t}\tilde\varphi_0(\vec{r}),
\end{equation}
with the initial condition $\tilde F_V(\vec{r},0)=0$. We can rewrite this inhomogeneous TDSE as an integral equation that takes the initial condition into account:
\begin{equation}
\label{eq_fv}
\tilde F_V(\vec{r},t)=\tilde F_V^{(1)}(\vec{r},t)+\mathrm{i}Z\int\frac{\mathrm{d}^3p}{(2\pi)^3}\tilde\chi_V(\vec{r},\vec{p},t)\int_0^t \mathrm{d}t'\;\int \frac{\mathrm{d}^3 r'}{r'}\tilde\chi^*_V(\vec{r}\;',\vec{p},t')\tilde F_V(\vec{r}\;',t'),
\end{equation}
with
\begin{equation}
\label{eq_fv1}
\tilde F_V^{(1)}(\vec{r},t)=\int\frac{\mathrm{d}^3p}{(2\pi)^3}\tilde\chi_V(\vec{r},\vec{p},t)\int_0^t \mathrm{d}t' \ \int \mathrm{d}^3 r' \tilde\chi^*_V(\vec{r}\;',\vec{p},t')
\left[b'(t')e^{-\mathrm{i}\varepsilon_0 t'}(\vec{e}\cdot\vec{\nabla_{r'}})\tilde\varphi_0(\vec{r'})-\mathrm{i}\zeta'(t')\right].
\end{equation}
We now consider the L-gauge and use the same ansatz:
\begin{equation}
\label{eq_phi_L2}
\tilde\Phi_L(\vec{r},t)=e^{-\mathrm{i}\varepsilon_0 t}\tilde\varphi_0(\vec{r})+\tilde F_L(\vec{r},t).
\end{equation}
Similar manipulations with Eq. (\ref{eq_tdse_lg})  lead to the following integral equation for $\tilde F_L(\vec{r},t)$:
\begin{equation}
\label{eq_fl}
\tilde F_L(\vec{r},t)=\tilde F_L^{(1)}(\vec{r},t)+\mathrm{i}Z\int\frac{\mathrm{d}^3p}{(2\pi)^3}\tilde\chi_L(\vec{r},\vec{p},t)\int_0^t \mathrm{d}t'\;\int \frac{\mathrm{d}^3 r'}{r'}\tilde\chi^*_L(\vec{r}\;',\vec{p},t')\tilde F_L(\vec{r}\;',t'),
\end{equation}
with
\begin{equation}
\label{eq_fl1}
\tilde F_L^{(1)}(\vec{r},t)=-\mathrm{i}\int\frac{\mathrm{d}^3p}{(2\pi)^3}\tilde\chi_L(\vec{r},\vec{p},t)\int_0^t \mathrm{d}t' b''(t')e^{-\mathrm{i}\varepsilon_0 t'}\;\int\mathrm{d}^3 r'
\tilde\chi^*_L(\vec{r}\;',\vec{p},t')(\vec{e}\cdot\vec{r}\;')\tilde\varphi_0(\vec{r'}).
\end{equation}
We easily see that even the free terms (\ref{eq_fv1}) and (\ref{eq_fl1}) are different. This implies that the perturbative (Born) terms generated from Eqs (\ref{eq_fv}) and (\ref{eq_fl}) are not gauge
invariant because our ansatz Eqs (\ref{eq_phi_V1}) and (\ref{eq_phi_L2}) do not satisfy the G\"oppert-Mayer transformation. In order to satisfy the gauge invariance, we have to change one of the ansatz. In the L-gauge, taking into account the G\"oppert-Mayer gauge transformation (\ref{eq_GM_transf}), we 
rewrite our ansatz as follows:
\begin{equation}
\tilde\Phi_L(\vec{r},t)=e^{-\mathrm{i}\varepsilon_0 t-\mathrm{i}b'(t)(\vec{e}\cdot\vec{r})}\tilde\varphi_0(\vec{r})+\tilde{\bar F}_L(\vec{r},t).
\end{equation}
Note that the introduction of a phase factor does not modify the physics.\\

The previous discussion shows that we have some freedom in the definition of the ansatz through the phase factor. Hereafter, we define two families of ansatz in such a way
that within a given family, both the L-gauge and the V-gauge versions of the ansatz lead to the same results. In other words, $\tilde\Phi_V$ and $\tilde\Phi_L$ as well as 
each term of the perturbation expansion of $\tilde F_V$ and $\tilde F_L$ satisfy relation (\ref{eq_GM_transf}).\\

In the first family, the V-gauge and L-gauge ansatz are defined as above.  The V-gauge ansatz is:
\begin{equation}
\label{eq_phi1_V}
\tilde\Phi_V(\vec{r},t)=e^{-\mathrm{i}\varepsilon_0 t}\tilde\varphi_0(\vec{r})+\tilde F_{1,V}(\vec{r},t),
\end{equation}
where $\tilde F_{1,V}(\vec{r},t)$ satisfies the inhomogeneous TDSE Eq. (\ref{eq_inh_tdse_vg}) (reproduced here for clarity):
\begin{equation}
\label{eq_inh_tdse_vg1}
\left[\mathrm{i}\frac{\partial}{\partial t}+\frac{1}{2}\triangle_r- \mathrm{i}b'(t)(\vec{e}\cdot\vec{\nabla}_r)+\frac{Z}{r}-\zeta'(t)\right] \tilde F_{1,V}(\vec{r},t) = [\mathrm{i}b'(t)
(\vec{e}\cdot\vec{\nabla}_r)+\zeta'(t)]e^{-\mathrm{i}\varepsilon_0 t}\tilde\varphi_0(\vec{r}),
\end{equation}
while the L-gauge ansatz is given by:
\begin{equation}
\label{eq_phi1_L}
\tilde\Phi_L(\vec{r},t)=e^{-\mathrm{i}b'(t)(\vec{e}\cdot\vec{r})-\mathrm{i}\varepsilon_0 t}\tilde\varphi_0(r)+\tilde F_{1,L}(\vec{r},t),
\end{equation}
where $\tilde F_{1,L}(\vec{r},t)$ satisfies the following inhomogeneous TDSE:
\begin{equation}
\label{eq_inh_tdse_lg1}
\left[\mathrm{i}\frac{\partial}{\partial t}+\frac{1}{2}\triangle_r- b''(t)(\vec{e}\cdot\vec{r})+\frac{Z}{r}\right] \tilde F_{1,L}(\vec{r},t) = e^{-\mathrm{i}b'(t)(\vec{e}\cdot\vec{r})}[\mathrm{i}b'(t)(\vec{e}\cdot\vec{\nabla}_r)+\zeta'(t)]e^{-\mathrm{i}\varepsilon_0 t}\tilde\varphi_0(\vec{r}).
\end{equation}

\vspace{1cm}
In the second family, we define the L-gauge ansatz as follows:
\begin{equation}
\label{eq_phi2_L}
\tilde\Phi_L(\vec{r},t)=e^{-\mathrm{i}\varepsilon_0 t}\tilde\varphi_0(\vec{r})+\tilde F_{2,L}(\vec{r},t),
\end{equation}
where $\tilde F_{2,L}(\vec{r},t)$ satisfies the following inhomogeneous TDSE:
\begin{equation}
\label{eq_inh_tdse_lg2}
\left[\mathrm{i}\frac{\partial}{\partial t}+\frac{1}{2}\triangle_r-b''(t)(\vec{e}\cdot\vec{r})+\frac{Z}{r}\right] \tilde F_{2,L}(\vec{r},t) = b''(t)e^{-\mathrm{i}\varepsilon_0 t}(\vec{e}\cdot\vec{r}){\tilde\varphi}_0(\vec{r}).
\end{equation}
In this case, the V-gauge ansatz is:
\begin{equation}
\label{eq_phi2_V}
\tilde\Phi_V(\vec{r},t)=e^{\mathrm{i}b'(t)(\vec{e}\cdot\vec{r})-\mathrm{i}\varepsilon_0 t}\tilde\varphi_0(\vec{r})+\tilde F_{2,V}(\vec{r},t),
\end{equation}
where $\tilde F_{2,V}(\vec{r},t)$ satisfies the following inhomogeneous TDSE:
\begin{equation}
\label{eq_inh_tdse_vg2}
\left[\mathrm{i}\frac{\partial}{\partial t}+\frac{1}{2}\triangle_r- \mathrm{i}b'(t)(\vec{e}\cdot\vec{\nabla}_r)+\frac{Z}{r}-\zeta'(t)\right] \tilde F_{2,V}(\vec{r},t) = b''(t)e^{\mathrm{i}b'(t)(\vec{e}\cdot\vec{r})
-\mathrm{i}\varepsilon_0 t}(\vec{e}\cdot\vec{r})\tilde\varphi_0(\vec{r}).
\end{equation}
Within a given family, it is easy to show that the gauge invariance is satisfied. Not only the ansatz are gauge invariant but also each term of the perturbation expansion of the full wave packet. However, if we compare, say the V-gauge ansatz of the two  families, the terms of the perturbation expansion of the full wave packet are different at each order, but the total sum of all these terms should remain the same. It is important to note that if we solve exactly the inhomogeneous TDSE associated to the ansatz we have introduced,  they must give the same result. The 
situation is different if we consider an approximate scheme such as the SFA. Usually, the SFA consists in keeping only the first term of the perturbation expansion (in the Coulomb potential) of the inhomogeneous TDSE. If more terms are taken into account, we get what we call a high order SFA which describes the multiple rescattering of the electron by the Coulomb potential of the 
ionic core. Within the present approach, we have therefore defined two SFA schemes that correspond to our two families of ansatz. It is now important to establish the connection between our
two SFA schemes and  the well known SFA schemes. It is easy to see that the V-gauge SFA as defined in \cite{choice} corresponds to our V-gauge ansatz of the first family 
while the L-gauge ansatz of the second family leads to PPT. Let us emphasize, that two wave functions in the same gauge but belonging to different families, {\it e.g.} given by Eqs (\ref{eq_phi1_V}) and (\ref{eq_phi2_V}), are equal. The sum of the two terms is the same independently of the ansatz, but what is different is the contribution of each term.

\section{Numerical treatment of the TDSE within SFA  }

\subsection{Semi-analytical formulae}

\subsubsection{First order SFA}

Within the SFA, it is often possible to derive semi-analytical expressions or more precisely, integral representations of the full wave packet as illustrated below. We first consider 
our V-gauge ansatz of the first family. In order to obtain simple expressions for the full wave packet in this case, it is more convenient to work in the momentum space. By means 
of Eqs (\ref{eq_phi_V1}) and (\ref{eq_fv1}), we obtain after some manipulations:
\begin{eqnarray}
\Phi_{1,V}^{SFA}(\vec{p},t)&=&e^{-\mathrm{i}\varepsilon_0 t}\varphi_0(\vec{p}) +F^{(1)}_{1,V}(\vec{p},t)\nonumber\\
&=&e^{-\mathrm{i}\varepsilon_0 t}\varphi_0(\vec{p})+ \mathrm{i}\varphi_0(\vec{p})e^{-\mathrm{i}S(\vec{p},t)}\int_0^t\ e^{\mathrm{i}S(\vec{p},\xi)-\mathrm{i}\varepsilon_0\xi}\left[b'(\xi)
(\vec{e}\cdot\vec{p})-\mathrm{i}\zeta'(\xi)\right] \ \mathrm{d}\xi \nonumber\\
&=&e^{-\mathrm{i}S(\vec{p},t)}\left[\varphi_0(\vec{p})+\mathrm{i}(p^2/2-\varepsilon_0)\varphi_0(\vec{p})\int_0^te^{\mathrm{i}S(\vec{p},\xi)-\mathrm{i}\varepsilon_0\xi}\mathrm{d}\xi\right].
\end{eqnarray}
We use this approximation of the full wave packet to calculate the electron energy spectrum along the polarization axis and the probability to stay in the 1s-state at a time $T$ corresponding to the end of the pulse. The electron energy spectrum is usually obtained by projecting the above wave packet on plane waves of wave vector $\vec{k}$ , having taken care to subtract from the wave packet, the contribution of the ground state which is not orthogonal to a  plane wave. The transition matrix element is:
\begin{eqnarray}
\langle\vec{k}e^{-\mathrm{i}k^2t/2}|\Phi_{1,V}^{SFA}(t)-e^{-\mathrm{i}\varepsilon_0 t}\varphi_0\rangle&=&\langle\vec{k}e^{-\mathrm{i}k^2t/2}|F^{(1)}_{1,V}(t)\rangle\nonumber\\
\label{eq_spec_sub_vg}
&=&e^{\mathrm{i}b(t)(\vec{e}\cdot\vec{k})-\mathrm{i}\zeta(t)}\varphi_0(\vec{k})\left[1+\mathrm{i}(k^2/2-\varepsilon_0)\int^t_0\mathrm{d}\xi e^{\mathrm{i}S(\vec{k},\xi)-\mathrm{i}\varepsilon_0\xi}\right]-
e^{\mathrm{i}(k^2/2-\varepsilon_0)t}\varphi_0(\vec{k}).
\end{eqnarray}
Within an irrelevant phase factor, this expression of the transition matrix element coincides with the expression given, in the V-gauge, by Bauer {\it et al.} in  \cite{choice}. However it is 
important to stress that the procedure of subtracting the ground state from the wave packet is only valid if the ground state component of function $F^{(1)}_{1,V}$ can be neglected. In other
words, we have to assume that the ground state is not depleted. This can be seen easily by projecting the full wave packet on modified plane waves, denoted by $\varphi_{\bot}(\vec{k})$, 
that are orthogonalized to the ground state. Using Dirac notation, we have:
\begin{equation}
\langle\varphi_{\bot}|=\langle\vec{k}|-\langle\vec{k}|\varphi_0\rangle\langle\varphi_0|.
\end{equation}
In this case, the transition matrix element becomes:
\begin{equation}
\langle\varphi_{\bot}e^{-\mathrm{i}k^2t/2}|\Phi_{1,V}^{SFA}(t)\rangle=\langle\vec{k}e^{-\mathrm{i}k^2t/2}|F_{1,V}^{(1)}(t)\rangle+e^{\mathrm{i}(k^2/2-\varepsilon_0)t}\varphi_0(\vec{k})[1-\langle\varphi_0
e^{-\mathrm{i}\varepsilon_0 t}|\Phi_{1,V}^{SFA}(t)\rangle].
\end{equation}
We clearly see that this last transition matrix element reduces to the expression (\ref{eq_spec_sub_vg}) if $\langle\varphi_0 e^{-\mathrm{i}\varepsilon_0 t}|\Phi_{1,V}^{SFA}(t)\rangle$, which represents the amplitude
to stay in the ground state at the end of the pulse,  tends to 1 or equivalently if the ground state is not depleted. Let us mention that a more rigorous calculation of the electron energy spectrum requires the projection of the full wave packet on Coulomb waves with incoming wave asymptotic behavior. This calculation is tremendously difficult and we address this problem in the second part of this section where we develop a fully numerical approach. Finally, we want to point out that the full wave packet should be normalized before calculating the spectrum. Our reformulation of the SFA that leads to the solution of an inhomogeneous TDSE clearly shows that the norm of the full wave packet is not conserved. On the other hand, in all studies where the validity of the SFA is assessed, the SFA results for the electron energy spectrum are normalized by forcing the latter to coincide with the TDSE results at a given energy. In other words, it may appear that the normalization factor is 
irrelevant. However, if we want to compare SFA results using an ansatz from different families or simply evaluate the probability to stay in the ground state at the end of the pulse, the normalization factor may become important. This factor is given by:
\begin{eqnarray}
\langle\Phi_{1,V}^{SFA}|\Phi_{1,V}^{SFA}\rangle&=&1+\frac{4\mathrm{i}}{\pi^2}\int\frac{\mathrm{d}^3p}{(p^2+1)^3}\int_0^te^{\mathrm{i}(S(\vec{p},\xi)-\varepsilon_0\xi)}\mathrm{d}\xi+c.c.\nonumber\\
\label{eq_norm_anal}
&+&\frac{2}{\pi^2}\int\frac{\mathrm{d}^3p}{(p^2+1)^2}\int_0^te^{-\mathrm{i}(S(\vec{p},\xi_1)-\varepsilon_0\xi_1)}\mathrm{d}\xi_1\int_0^t e^{\mathrm{i}(S(\vec{p},\xi_2)-\varepsilon_0\xi_2)}\mathrm{d}\xi_2,
\end{eqnarray}
where $c.c.$ stands for 'complex conjugate'. Let us note that in the limit where the ground state is not depleted, this norm goes to 1. If, on the contrary, the ground state is partially depleted,  the probability amplitude to stay in the ground state is calculated by projecting the full wave packet on the ground state,
\begin{equation}
\langle\varphi_0e^{-\mathrm{i}\varepsilon_0t}|\Phi_{1,V}^{SFA}\rangle=\frac{8}{\pi^2}\int\frac{\mathrm{d}^3p}{(p^2+1)^4}e^{-\mathrm{i}(S(\vec{p},t)-\varepsilon_0t)}+\frac{4\mathrm{i}}{\pi^2}\int\frac{\mathrm{d}^3p}{(p^2+1)^3}\int_0^te^{-\mathrm{i}(S(\vec{p},t)-S(\vec{p},\xi))+\mathrm{i}\varepsilon_0(t-\xi)}\mathrm{d}\xi,
\end{equation}
and by dividing the result by the square root of the norm (\ref{eq_norm_anal}).\\

Let us now consider the expression of the full wave packet $\tilde\Phi_{2,L}(\vec{r},t)$ that corresponds to our L-gauge ansatz of the second family. Starting from Eqs (\ref{eq_phi2_L}) and (\ref{eq_inh_tdse_lg2}), we obtain:
\begin{eqnarray}
\tilde\Phi^{SFA}_{2,L}(\vec{r},t)&=&e^{-\mathrm{i}\varepsilon_0 t}\tilde\varphi_0(\vec{r}) - \mathrm{i}\int\frac{\mathrm{d}^3p}{(2\pi)^3}\tilde\chi_L(\vec{r},\vec{p},t)\int_0^t\mathrm{d}\xi\ b''(\xi)
e^{-\mathrm{i}\varepsilon_0\xi}\int \mathrm{d}^3 r' \tilde\chi^*_L(\vec{r}\;',\vec{p},\xi)(\vec{e}\cdot\vec{r}\;')\tilde\varphi_0(\vec{r}\;') \nonumber\\
\label{eq_phiL_sfa}
&=&e^{-\mathrm{i}b'(t)(\vec{e}\cdot\vec{r})}\int\frac{\mathrm{d}^3p}{(2\pi)^3}\ e^{\mathrm{i}\vec{p}\cdot\vec{r}}\Phi^{SFA}_{2,V}(\vec{p},t)
\end{eqnarray}
where
\begin{equation}
\Phi^{SFA}_{2,V}(\vec{p},t)=e^{-\mathrm{i}S(\vec{p},t)}\left[\varphi_0(\vec{p})+\mathrm{i}\int_0^t\ d\xi[P^2(\xi)/2-\varepsilon_0]\varphi_0(\vec{P}(\xi)) e^{\mathrm{i}S(\vec{p},\xi)-\mathrm{i}\varepsilon_0\xi}\right].
\end{equation}
This last expression follows from Eqs (\ref{eq_phi2_V}) and (\ref{eq_inh_tdse_vg2}). Note that expression (\ref{eq_phiL_sfa}) for the full wave packet leads to PPT's result. In addition, projecting $\tilde\Phi_L^{SFA}$ on plane (Volkov) waves
gives Keldysh transition matrix element within a phase factor. The electron energy spectra as well as the probability to stay in the ground state at the end of the pulse can be calculated in the
same way as before.\\

\subsubsection{First and second order SFA}

In the tunneling regime, the numerical calculation of the above semi-analytical expressions that involves multiple integrations of highly oscillating functions is extremely difficult thereby requiring special techniques such as the stationary phase method. Needless to say that high order terms in the Coulomb potential that describe multiple re-scattering of the electrons by the ionic core are even more complicated to calculate. In addition to multiple integrations, the Coulomb potential introduces singularities that have to be treated with care. The forward re-scattering of slow returning electrons by the ionic core has been invoked  by several groups \cite{becker_les,milosevic_les,guo} to explain the existence of the famous low and very low energy structures in the energy distribution of the emitted electrons along the polarization axis \cite{blaga,quan,wu}. They calculated the second order term namely the so-called re-scattering term within the SFA while treating the singularity (related to the divergence of the Rutherford scattering in the forward direction)  in a rather heuristic way. Very recently however, Titi and Drake \cite{titi} presented an accurate treatment of this singularity by introducing a regularization scheme in the case of an electric field of constant intensity. \\

For the sake of illustration and in order to get more insight into the role of the Coulomb potential, we calculate here the first two terms of the expansion of the SFA amplitude in power of the Coulomb potential in the case of the interaction of atomic hydrogen with a laser pulse, the vector potential of which has the following form:
\begin{equation}
\label{eq_sin_pulse}
b'(t)=\frac{1}{\omega_0} \sqrt{\frac{I}{I_0}}\sin^2(\pi\frac{t}{T})\sin(\omega_0t+\phi).
\end{equation}
Here, $\omega_0=0.057$ a.u. is the field frequency and $T=2\pi N/\omega_0$ is the total pulse duration with $N=6$, the total number of optical cycles. $I=8.7$ 10$^{13}$ Watt/cm$^2$ is the peak intensity, $I_0=3.5$ 10$^{16}$ Watt/cm$^2$ being the atomic unit of intensity. $\phi$ is the carrier phase which is set equal to zero in the following. Our calculations rely on several approximations that are discussed in the following. We start from the L-gauge ansatz (32) of the second family and write the solution $\tilde F_{2,L}(\vec{r},t)$ of  the inhomogeneous TDSE as an expansion in terms of plane waves:
\begin{equation}
\tilde F_{2,L}(\vec{r},t)=\int\frac{\mathrm{d}\vec{k}}{(2\pi)^3}\;M(\vec{k},t)\langle\vec{r}|\vec{k}\rangle.
\end{equation}
Using this condition and if we assume that there is no depletion of the ground state which means that the norm of the total wave packet is equal to one, the present formulation is very close to the so-called Lewenstein model \cite{lewenstein_rings}. Note however that in Lewenstein's model, $\tilde F_{2,L}(\vec{r},t)$ is actually expanded on the eigenstates of the field free Hamiltonian $i.e$ in Coulomb waves.  After substituting ansatz (32) into the TDSE and taking into account expansion (45) we obtain the following equation for the amplitude $M(\vec{k},t)$:   
\begin{equation}
\label{eq_b_PW}
\frac{\partial}{\partial t}M(\vec{k},t) = -\mathrm{i}(\frac{k^2}{2}-\varepsilon_0)M(\vec{k},t)-\mathrm{i}b''(t)d_z(\vec{k})+b''(t)\frac{\partial M(\vec{k},t)}{\partial k_z} + \int \frac{d\vec{u}}{(2\pi)^3} \; M(\vec{u},t) \langle \vec{k} |V| \vec{u} \rangle,
\end{equation}
where $d_z(\vec{k})=\langle \vec{k} | z | \tilde\varphi_0 \rangle$ is the dipole coupling between the ground state and a continuum wave function (here the plane wave $|\vec{k}\rangle$). $V$, for the time being, is the Coulomb potential.  Using the expansion in terms of field free Hamiltonian eigenstates leads to  the Eq. (10) of ref. \cite{lewenstein_rings}.
 Eq. (\ref{eq_b_PW}) has the great advantage of decoupling the contribution of the potential $V$ in the wave packet dynamics while in \cite{lewenstein_rings}, the Coulomb effect is introduced by using Coulomb waves for the field free Hamiltonian eigenstates. In proceeding in this way, the orthogonality of the Coulomb waves to the ground state $\varphi_0$ is preserved. The difficulty of Lewenstein's formalism lies therefore in a proper description of the dipole transitions between Coulomb waves. In \cite{lewenstein_rings}, this dipole element is approximated as $\mathrm{i}\nabla_{k_z}\delta(\vec{k}-\vec{u})+g(\vec{k},\vec{u})$. One recognizes in the first term the contribution of the plane wave asymptotic behavior of the Coulomb wave, while the $g$ term includes the scattering amplitude. Neglecting the effect of $V$ in Eq. (\ref{eq_b_PW}) and setting $g=0$ in Eq. (10) of ref.\cite{lewenstein_rings}  gives the same following result:
\begin{equation}
\label{eq_b0}
M^{(1)}(\vec{p},T) \equiv F_{2,L}^{(1)}(\vec{p},T)=-\mathrm{i}e^{-\mathrm{i}S(\vec{p},T)} \int_0^T \mathrm{d}\xi \; b''(\xi)d_z(\vec{P}(\xi))e^{\mathrm{i}(S(\vec{p},\xi)-\varepsilon_0\xi)} ,
\end{equation} 
 which is the lowest order perturbation as described in the previous sections. The photoelectron spectrum is given by $p^2|M(\vec{p})|^2$. As mentioned in the previous sections,  we should, in principle, project the wave function onto exact eigenstates of the field free Hamiltonian. But for the sake of simplicity, plane waves will be used in this semi-analytical approximation. The next order can also be calculated analytically and the solution is:
\begin{equation}
\label{eq_b1}
M^{(2)}(\vec{p},T) \equiv F_{2,L}^{(2)}(\vec{p},T)=-e^{-\mathrm{i}S(\vec{p},T)}\int_0^T \mathrm{d}\xi\; e^{\mathrm{i}S(\vec{p},\xi)} \; \int_0^{\xi} \mathrm{d}\eta\;b''(\eta)e^{-\mathrm{i}\varepsilon_0\eta} \; \int \frac{\mathrm{d}\vec{u}}{(2\pi)^3} \; \langle \vec{p} |V| \vec{u} \rangle e^{-\mathrm{i}S(\vec{u},\xi)+\mathrm{i}S(\vec{u},\eta)} d_z(\vec{U}(\xi)) .
\end{equation}
where $\vec{U}(\xi)$ is related to $\vec{u}$ by relation (4). Note the presence of the coupling term $\langle \vec{p} |V| \vec{u} \rangle = -\frac{4\pi}{|\vec{p}-\vec{u}|^2}$ for the atomic hydrogen case. The $M^{(1)}$ and $M^{(2)}$ terms have also been introduced in \cite{milosevic2006} using the S-matrix formalism and the completness of the Volkov basis. There is a phase difference between our results and those from \cite{milosevic2006}, which appears to be the same for $M^{(1)}$ and $M^{(2)}$ and thus plays no role. A singularity is present in the integrand of expression (48). This singularity is usually overcome by replacing the Coulomb potential by a Yukawa potential ($V(r)=\frac{e^{-\alpha r}}{r}$). 
\begin{figure}[h]
\includegraphics[scale=0.22]{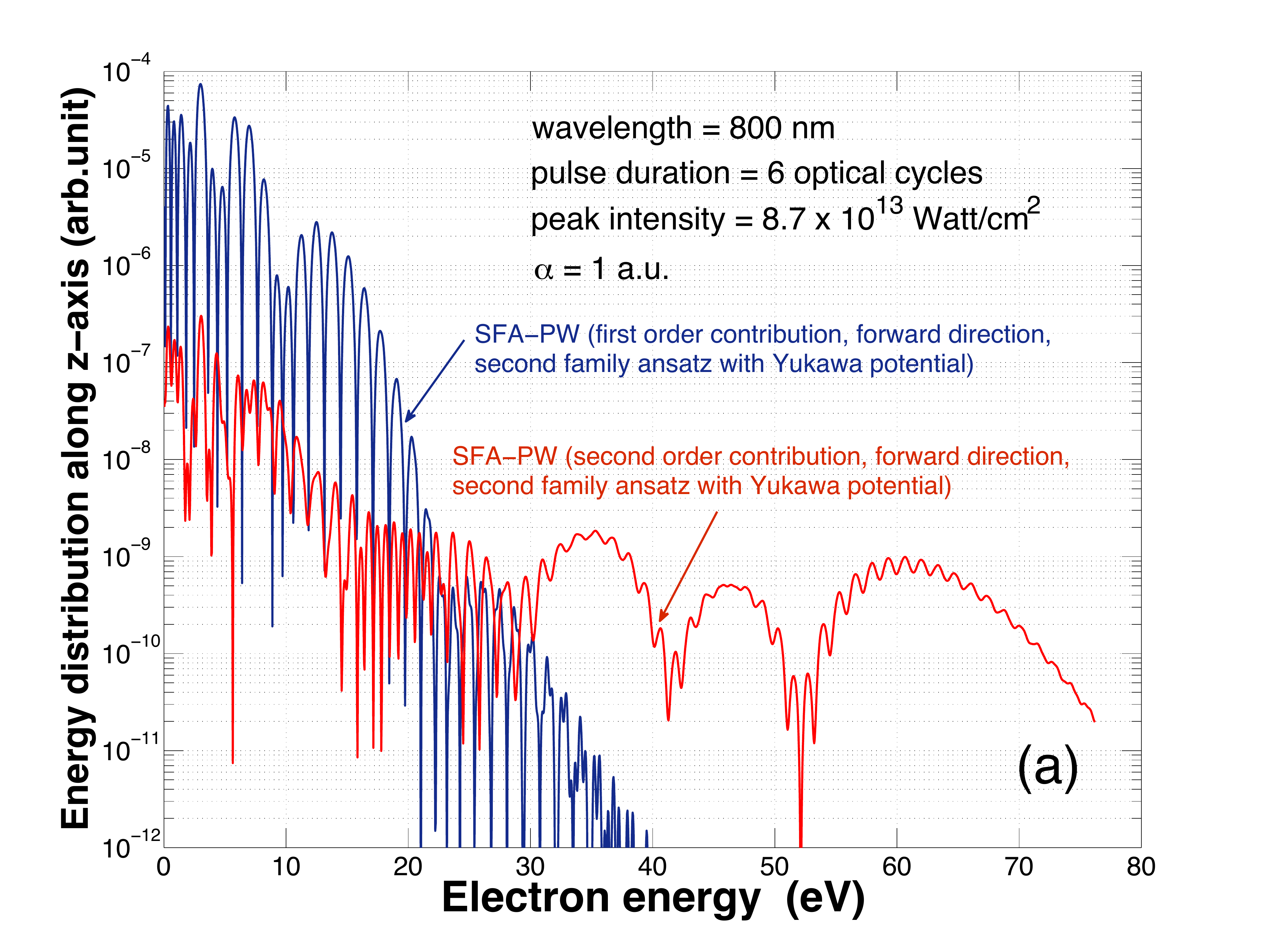}\includegraphics[scale=0.22]{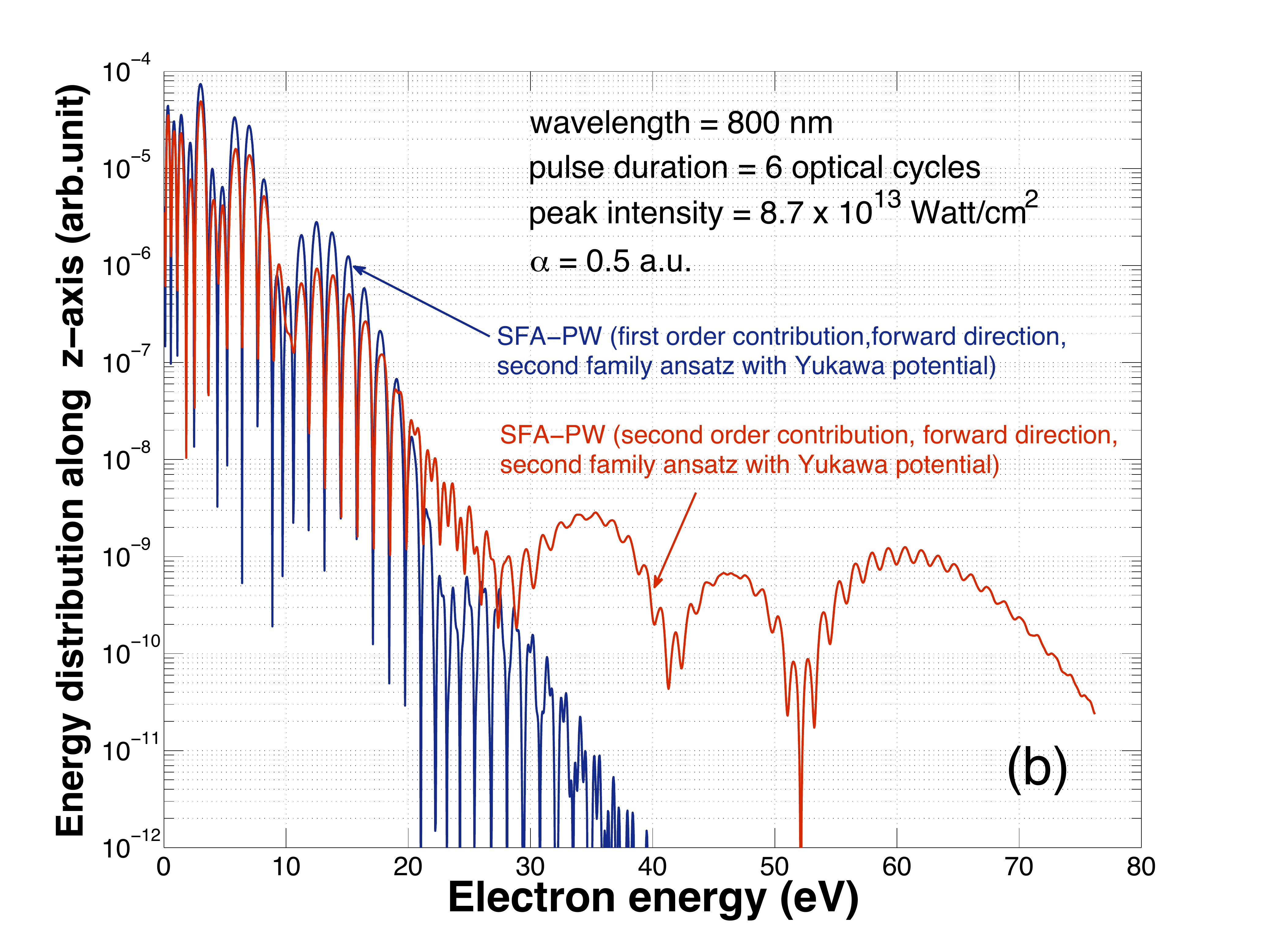}
\caption{(Color online) Energy distribution of the electrons emitted in the forward direction along the polarization axis and resulting from the interaction of atomic hydrogen with a 6-cycle pulse of  
                 frequency $\omega_0=0.057$ a.u. and peak intensity $I=8.7\times 10^{13}$ Watt/cm$^2$. The results are obtained within the SFA approximation by using the L-gauge ansatz of the 
                 second family. The final wave packet is projected on plane waves and the Coulomb potential replaced by a Yukawa potential of parameter $\alpha$. The blue curve corresponds to the 
                 modulus square of the first term of the expansion of the SFA amplitude in power of the Yukawa potential while the red curve corresponds to the second term of the expansion. Two 
                 values of $\alpha$ are considered: (a) $\alpha=1$ a.u. and (b) $\alpha=0.5$ a.u.}
\label{fig1}
\end{figure}
In Fig.\ref{fig1} we plot the result obtained for the energy distributions $|M^{(1)}|^2$, $|M^{(2)}|^2$ along the polarization axis for an ejection angle of $\theta = 0^{\circ}$. The integration over the intermediate momentum is performed by using the saddle point method. The contribution of the lowest order amplitude $M^{(1)}$ which describes direct electrons, is dominant for electron energies below $2U_p$ where $U_p$ is the ponderomotive potential given by $I/4\omega_0^2$ in atomic units. On the contrary, the second order term dominates in the range $2U_p - 10U_p$ where elastic re-collision occurs \cite{lewenstein_rings,Colosimo2008}. When the Yukawa potential parameter $\alpha$ is reduced from $1$ a.u. to $0.5$ a.u., $|M^{(2)}|^2$ becomes of the same magnitude as $|M^{(1)}|^2$. For energies below $2U_p$, $|M^{(2)}|^2$ actually starts to behave like $|M^{(1)}|^2$ while it is barely affected in the $2U_p - 10 U_p$ energy range. When $\alpha$ is even smaller, it starts to diverge below $2U_p$ due to the saddle point approximation and the divergence present in the integral. The integrand is diverging but it is an integrable divergence.  In conclusion, the term $|M^{(2)}|$ is larger than the $|M^{(1)}|$ and exhibits the same behavior in energy below $2U_p$ and the Coulomb asymptotic behavior manifests itself for low energy electrons as expected. Moreover, in Eq. (\ref{eq_b0}), this problem already arises when using the proper definition of $d_z$ if the wave function is expanded on eigenstates of the field free Hamiltonian, $i.e.$ on Coulomb waves as expressed in \cite{lewenstein_rings}. The dipole is then written as follows:          
\begin{equation}
\label{eq_dzcoul}
\langle\tilde{\varphi}^-(\vec{k})|z|\tilde{\varphi}_0\rangle= -\frac{8\sqrt(2)}{\pi}\mathrm{i}\frac{k_z}{(1+k^2)^3}(1+\mathrm{i}\gamma)\Gamma(1+\mathrm{i}\gamma)(1+\frac{k\gamma}{2})e^{2\phi\gamma}e^{-\pi\gamma/2} ,
\end{equation} 
with $\gamma = -\frac{Z}{k}$ and $\phi=\arctan(k)$. $\tilde{\varphi}^-(\vec{k},\vec{r})$ is a Coulomb wave with an asymptotic incoming wave behavior. When the argument of the dipole is close to zero, then $|\langle\tilde{\varphi}^-(\vec{k})|z|\tilde{\varphi}_0\rangle| \approx 1/\sqrt{k}$. In other words when integrating over $t$ in Eq.(\ref{eq_b0}) there are times for which $\vec{p} \approx b'(t)\vec{e}$ and $M^{(1)}$ can no longer be calculated with the formulation given in \cite{lewenstein_rings}. Note that this condition is sensitive to the angle of ejection. Along the polarization axis, the latter condition is not anymore  satisfied for energies above $2U_p$ so that the calculation of the integral in Eq.(\ref{eq_b0}) can be performed.

\subsection{Fully numerical treatment}
As stressed in the previous subsection, the calculation of the electron energy spectra as well as the probability to stay in the ground state at the end of the pulse by means of the above semi-analytical formulae is rather cumbersome and requires several approximations. By contrast, our reformulation of SFA offers the possibility of a fully numerical treatment allowing one to address unsolved questions such as the convergence of the perturbation expansion in the Coulomb potential of the full wave packet. In the following, we consider the calculation of the full wave packet from the V-gauge ansatz of the first family. We recall that this wave packet is:
\begin{equation}
\label{eq_phi_num_vg}
\tilde\Phi_V(\vec r,t)=e^{-\mathrm{i}\varepsilon_0 t}\tilde\varphi_0(\vec{r})+\tilde F_{1,V}(\vec{r},t),
\end{equation}
where the function $\tilde F_{1,V}(\vec{r},t)$ is the solution of the following inhomogeneous TDSE:
\begin{equation}
\label{eq_tdse_num_vg}
\left[\mathrm{i}\frac{\partial}{\partial t}+\frac{1}{2}\triangle_r- \mathrm{i}b'(t)(\vec{e}\cdot\vec{\nabla}_r)-\zeta'(t)\right] \tilde F_{1,V}(\vec{r},t)+\frac{Z}{r}\tilde F_{1,V}(\vec{r},t)=\left[\mathrm{i}b'(t)(\vec{e}\cdot\vec{\nabla}_r)+\zeta'(t)\right]e^{-\mathrm{i}\varepsilon_0 t}\tilde\varphi_0(\vec{r}).
\end{equation}
The right hand side term of this equation is a source term that leads to a transfer of population from the ground state to both bound and continuum s- and p-states. We solve this inhomogeneous equation by means of a spectral method which consists in expanding the solution in a basis of Coulomb sturmian functions \cite{rotenberg}. Note that all the time propagation methods such as the split operator method and the Crank-Nicolson method that are based on the well known expression of the time evolution operator in terms of the exponential of the Hamiltonian cannot be used directly. In the present case, we use a diagonally implicit Runge-Kutta method of order two \cite{alexander_dirk}. We have checked that if we solve Eq. (\ref{eq_tdse_num_vg}) without approximation and substitute the solution in the ansatz (\ref{eq_phi_num_vg}) to get the full wave packet, this latter one, properly normalized, coincides with the exact solution of the corresponding TDSE. Let us now solve the inhomogeneous equation (\ref{eq_tdse_num_vg}) iteratively to generate a perturbation expansion (Born series) of $\tilde F_{1,V}(\vec{r},t)$ in powers of the Coulomb potential. We write:
\begin{equation}
\label{eq_sum_F}
\tilde F_{1,V}(\vec{r},t)=\sum_{n=1}\tilde F_{1,V}^{(n)}(\vec{r},t).
\end{equation}
where $\tilde F_{1,V}^{(1)}(\vec{r},t)$ satisfies the following equation:
\begin{equation}
\label{eq_tdse_num_vg_o1}
\left[\mathrm{i}\frac{\partial}{\partial t}+\frac{1}{2}\triangle_r- \mathrm{i}b'(t)(\vec{e}\cdot\vec{\nabla}_r)-\zeta'(t)\right] \tilde F_{1,V}^{(1)}(\vec{r},t)=\left[\mathrm{i}b'(t)(\vec{e}\cdot\vec{\nabla}_r)+\zeta'(t)\right]e^{-\mathrm{i}\varepsilon_0 t}\tilde\varphi_0(\vec{r}),
\end{equation}
and $\tilde F_{1,V}^{(n)}(\vec{r},t)$ for $n>1$, the equation:
\begin{equation}
\label{eq_tdse_num_vg_on}
\left[\mathrm{i}\frac{\partial}{\partial t}+\frac{1}{2}\triangle_r- \mathrm{i}b'(t)(\vec{e}\cdot\vec{\nabla}_r)-\zeta'(t)\right] \tilde F_{1,V}^{(n)}(\vec{r},t)=-\frac{Z}{r}\tilde F_{1,V}^{(n-1)}(\vec{r},t).
\end{equation}
We note that at the end of the propagation, we normalize the wave function by dividing by the factor
\begin{equation}
\label{eq_norm_num}
N=\sqrt{\langle\tilde\varphi_0(t)+\tilde F_{1,V}(t)|\tilde\varphi_0(t)+\tilde F_{1,V}(t)\rangle}
\end{equation}
with $\tilde F_{1,V}(t)$ given by Eq. (\ref{eq_sum_F}).
The (usual) SFA consists in keeping only the first term $\tilde F_{1,V}^{(1)}(\vec{r},t)$ in the perturbation expansion (\ref{eq_sum_F}) while SFA of order $n$ with $n>1$ is obtained by keeping  the first $n$ terms of this 
expansion (\ref{eq_sum_F}). \\

\begin{figure}[h]
\includegraphics[width=12cm,height=8.5cm]{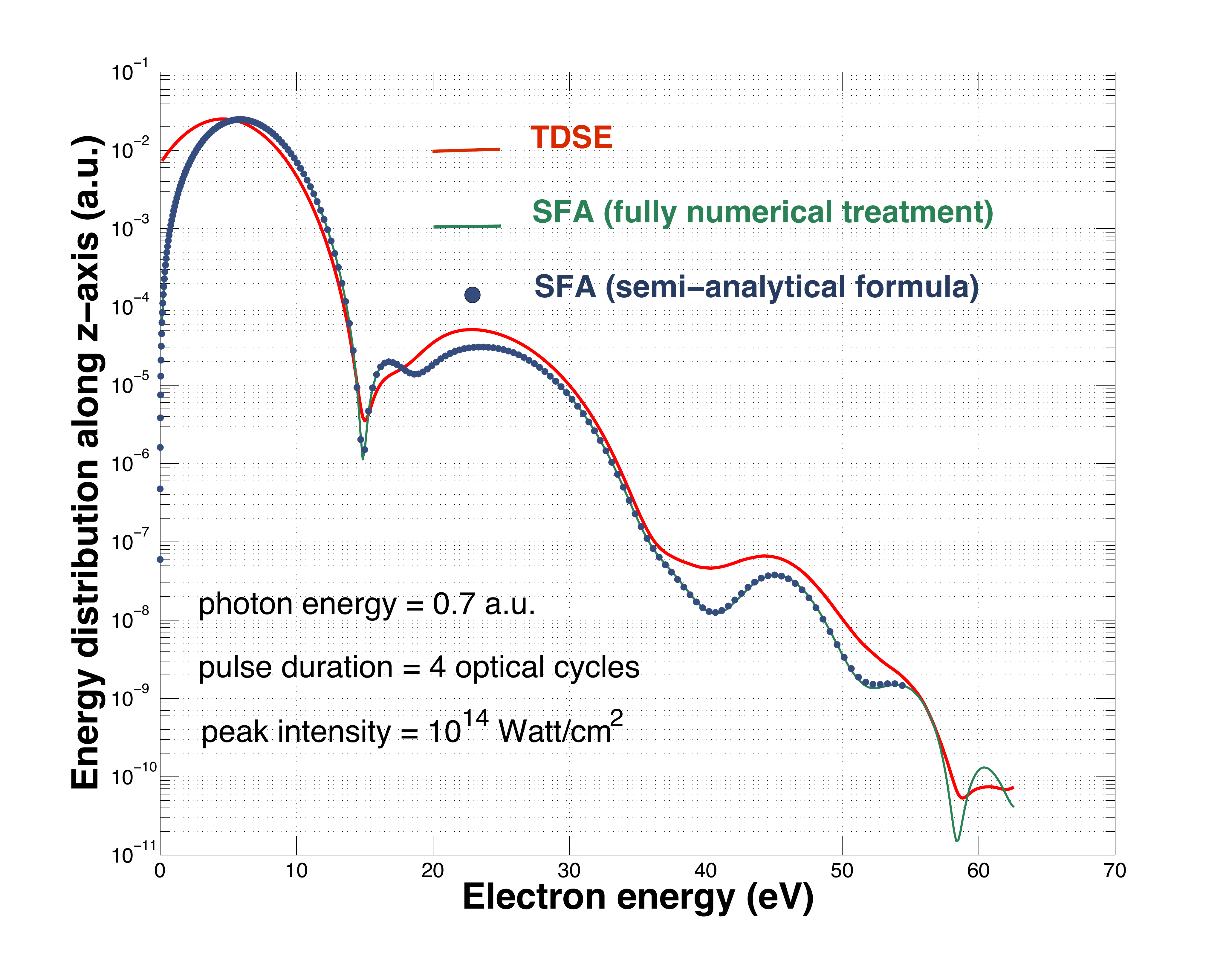}
\caption{(Color online) Energy distribution of the electrons emitted along the polarization axis, resulting from the interaction of atomic hydrogen with a 4-cycle pulse of frequency $\omega_0= 0.7$ 
                a.u. and peak intensity $I=10^{14}$ Watt/cm$^2$. The red curve corresponds to the TDSE results projected on the Coulomb wave. They are  compared to SFA results (in first order) obtained 
                by means of the V-gauge ansatz of the first family. The blue dots correspond to a semi-analytical calculation based on Eq. (\ref{eq_spec_sub_vg}) and the green curve corresponds to the 
                fully numerical treatment.}
\label{fig2}
\end{figure}
In order to show the pertinence of this fully numerical treatment and to study more in depth the validity of the SFA, let us consider a particularly simple case. As before, we assume that the vector potential is given by the formula (\ref{eq_sin_pulse}). In all the results presented in this section, we choose $\omega_0=0.7$ a.u., $N=4$, $I=10^{14}$ Watt/cm$^2$ and the carrier phase $\phi$ equal to zero. In Fig. \ref{fig2}, we show the energy distribution of the emitted electrons along the polarization axis. The red curve corresponds to the results obtained by solving the TDSE. We compare these TDSE results to those we obtain within SFA by using the V-gauge ansatz of the first family. The green curve and the blue dots correspond to a semi-analytical and a fully numerical calculation respectively. The semi-analytical calculations are based on formula (\ref{eq_spec_sub_vg}) divided by the norm (\ref{eq_norm_num}). Note that, in the present case, the norm is very close to one thereby justifying our way of removing the 1s state from the total wave 
packet. The fully numerical treatment consists in solving numerically Eq. (\ref{eq_tdse_num_vg_o1}). It is then necessary to normalize the full wave packet (\ref{eq_phi_num_vg}) and to remove from it the 1s state component before projecting on plane waves. We see in Fig. \ref{fig2} that the agreement between the two SFA calculations is perfect. However, it is important to note that, within SFA and contrary to what is usually believed, the population of the bound p-states is rather significant at the end of the interaction. Since the plane waves are not orthogonal to these p-states, they should be removed from the final SFA wave packet before projecting on the plane waves. The removal of these p-states affects very significantly the electron energy distribution for the whole range of energies considered here and destroys the rather good agreement we observe between SFA and TDSE results.\\

\begin{figure}[h]
\includegraphics[width=14cm,height=10cm]{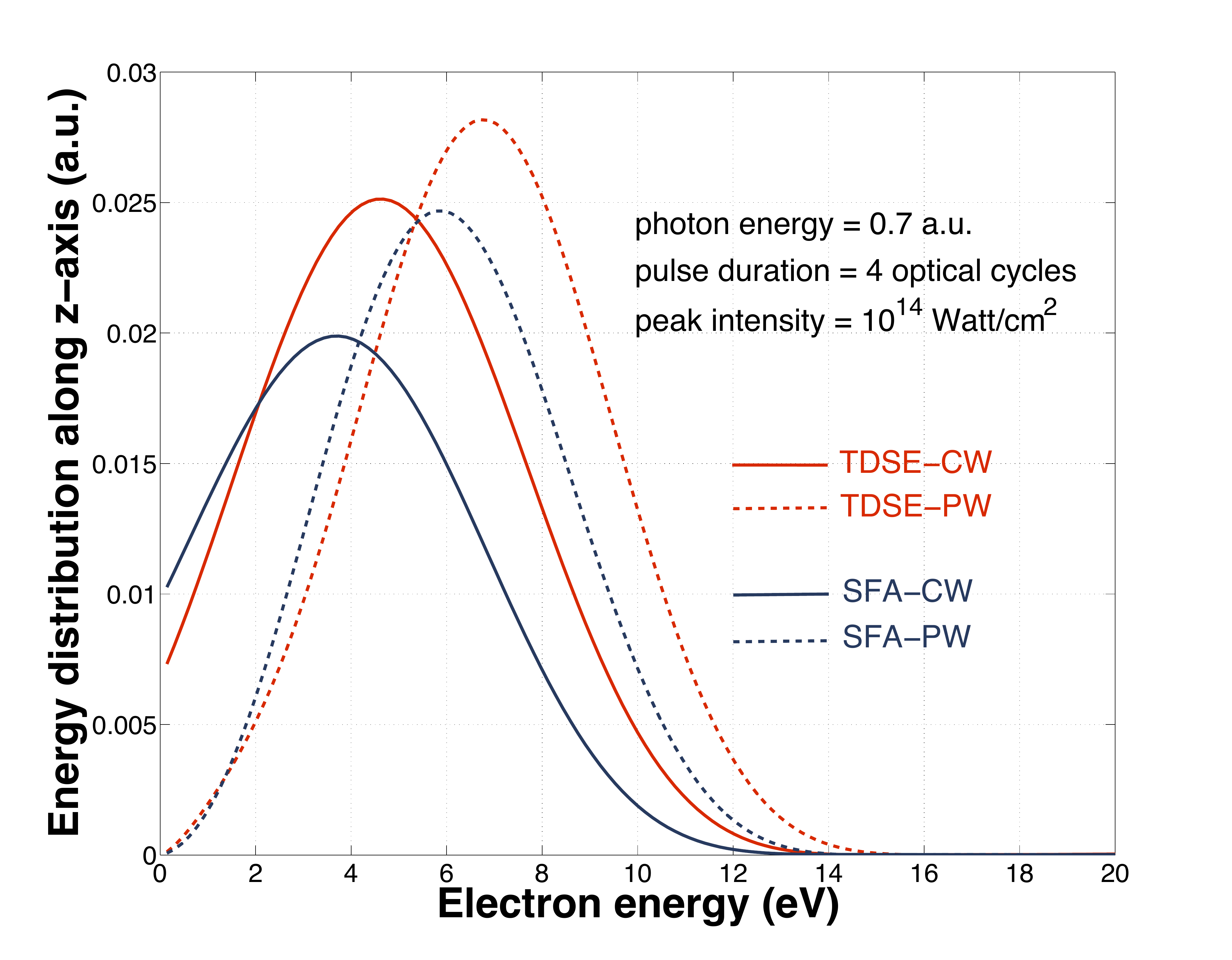}
\caption{(Color online) Energy distribution of the electrons emitted along the polarization axis, resulting from the interaction of atomic hydrogen with a 4-cycle pulse of frequency $\omega_0=0.7$
                a.u. and peak intensity $I=10^{14}$ Watt/cm$^2$. The red curves correspond to the TDSE results obtained by projecting the final wave packet on Coulomb waves (full line labelled 
                TDSE-CW) and plane waves (dashed line labelled TDSE-PW). The dark blue curves correspond to SFA results (in first order) obtained by projecting the final wave packet on Coulomb 
                waves (full line labelled SFA-CW) and on plane waves (dashed line labelled SFA-PW). }
\label{fig3}
\end{figure}

Let us now compare TDSE with SFA results  for the energy distribution obtained by projecting the final wave packet on Coulomb and plane waves. As shown in Fig. \ref{fig3}, the most significant  differences manifest in the low energy region of the energy distribution {\it i.e.} in the main peak in the present case. The SFA peak is shifted towards the right, simply  because the energy is not conserved when the full wave packet at the end of the pulse is projected on plane waves. In addition, due to the normalization factor of the plane waves, the SFA energy distribution is zero at zero electron energy. These conclusions are confirmed in Fig. \ref{fig3} where we compare TDSE and SFA results obtained by projecting the final wave packet on plane waves and Coulomb waves with an incoming wave character asymptotically. It is important to note that when the SFA wave packet is projected on Coulomb waves, the main peak is now shifted towards the left, {\it i.e.} towards the low energies. Furthermore, 
we also observe a decrease of the amplitude of this peak. This last feature is also true in the case of the TDSE results.

Our reformulation of the SFA allows one to calculate high order terms in the Coulomb potential and therefore to address the unsolved question of the convergence of the perturbation 
expansion (\ref{eq_sum_F}) \cite{anatomy}. As mentioned earlier, the second order term usually called the re-scattering term has been calculated (without normalization of the full wave packet) by 
several groups \cite{milosevic_rescattering,becker_rescattering,lewenstein_rings,bao_1996,lohr_1997,milosevic_1,milosevic_2} in the low frequency regime. These calculations show that back-scattering of the electrons by the ionic core is responsible for a plateau extending from about $2U_p$ to $10U_p$ in the energy spectrum. In addition, it has been proposed that the second order
term plays an important role in explaining 
\begin{figure}[h]
\includegraphics[width=12cm,height=8cm]{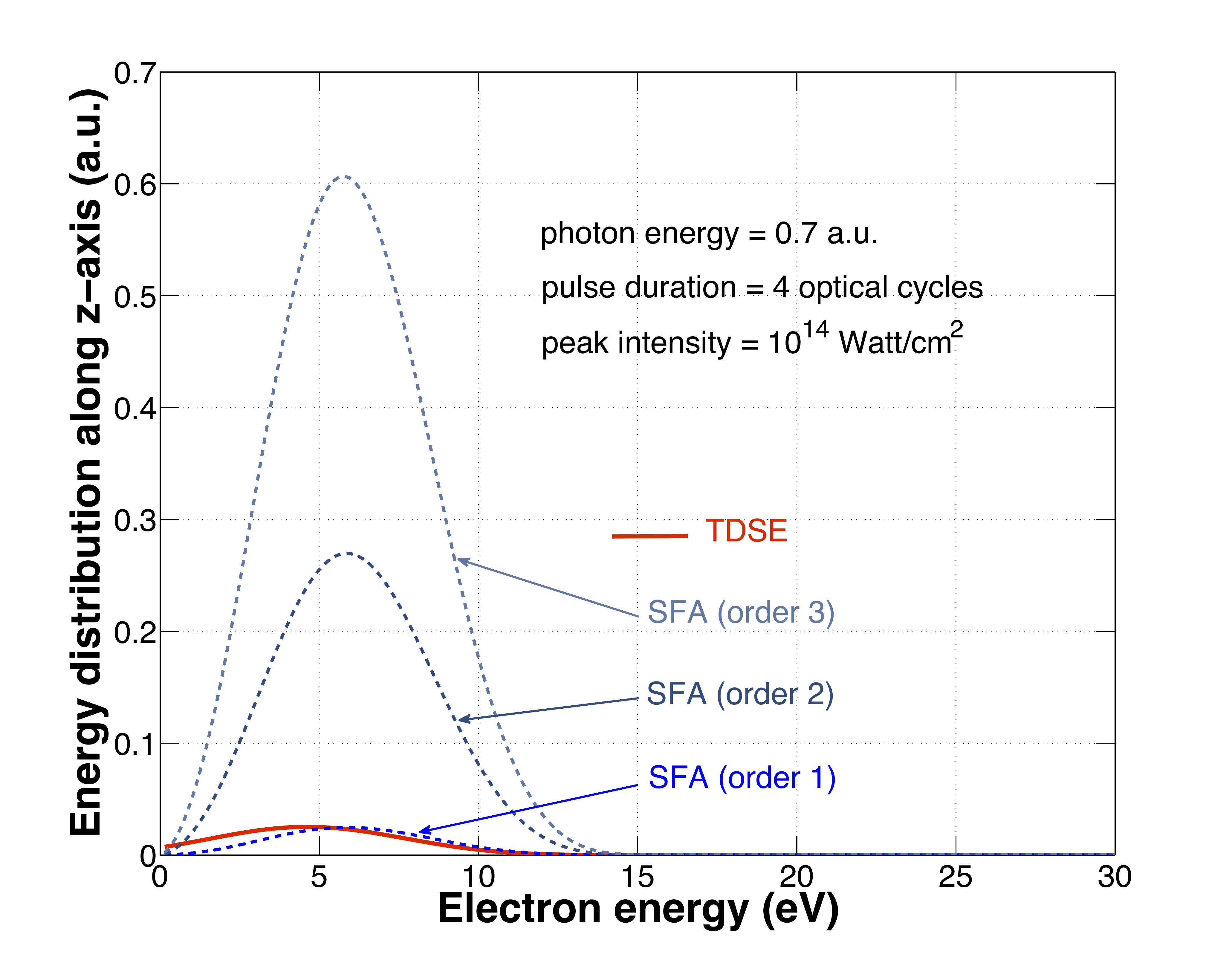}
\caption{(Color online) Energy distribution of the electrons emitted along the polarization axis, resulting from the interaction of atomic hydrogen with a 4-cycle pulse of frequency $\omega_0=0.7$
                a.u. and peak intensity $I=10^{14}$ Watt/cm$^2$. The red full curve corresponds to the TDSE results. The dashed curves correspond to the SFA results: the first, second and third
                orders are shown. }
\label{fig4}
\end{figure}
\begin{figure}[h]
\includegraphics[width=12cm,height=8cm]{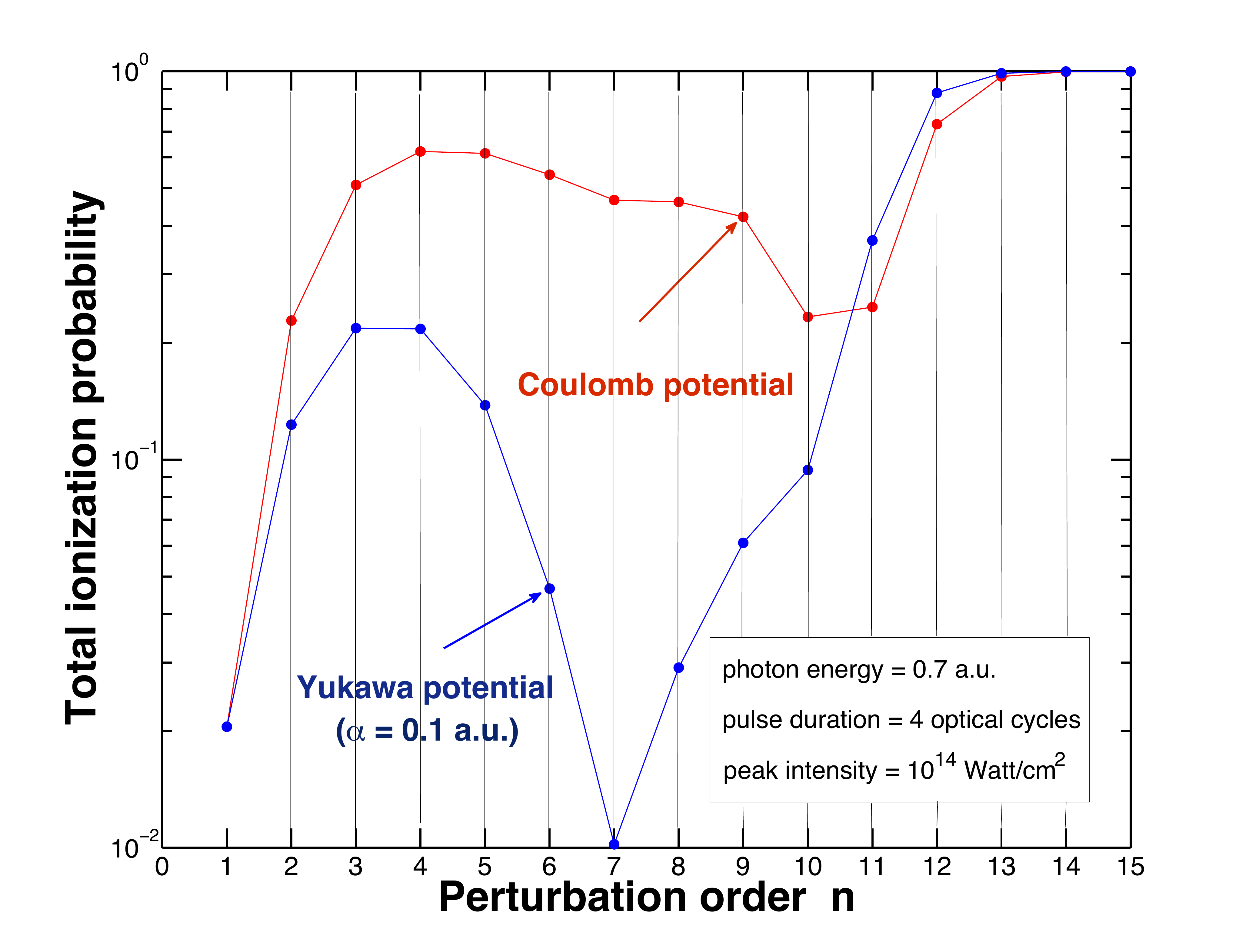}
\caption{(Color online) Total probability of ionization of atomic hydrogen interacting with a 4-cycle pulse of frequency $\omega_0=0.7$ a.u. and peak intensity $I=10^{14}$ Watt/cm$^2$ as a
                function of the order $n$ of the perturbation expansion of the full SFA wave packet. The results have been obtained for a pure Coulomb potential (red dots) and a Yukawa potential (blue
                dots). The parameter of the Yukawa potential is equal to 0.1 a.u.}
\label{fig5}
\end{figure}
the existence of the famous low energy structures \cite{blaga} as the result of a forward scattering of the returning electron by the ionic core \cite{becker_les,titi,milosevic_les}.  In the present case and instead of using semi-analytical formulae which turn out to be tremendously difficult to evaluate, we calculate high order terms by solving Eq. (\ref{eq_tdse_num_vg}) iteratively. Once this is done, we normalize the full SFA wave packet before projecting it on plane waves to get the electron energy distribution along the polarization axis. In Fig. \ref{fig4}, we compare our TDSE results to SFA results that include up to the third order corrections. It turns out that the second order SFA result that requires the calculation of $\tilde F_{1,V}^{(2)}(\vec{r},t)$, is about 10 times higher than the first order SFA result  that corresponds to 
the regular SFA. The third order term is still about twice the second order term. It is worth mentioning that we noticed that the SFA of the second order can be obtained from the first order simply by multiplying by a constant factor within a broad range of energies. Obviously, this raises the question of the convergence of the perturbation expansion of the full 
SFA wave packet. In order to try to answer to this question in this case, we show in Fig. \ref{fig5}, the total ionization probability as a function of the order $n$ of the perturbation expansion of the full SFA wave packet. We first perform the calculations for the Coulomb potential and then replace the Coulomb potential by the Yukawa potential $-Z\exp(-r/10)/r$ in Eq. (\ref{eq_tdse_num_vg_on}). In the case of the Coulomb potential we see that the total ionization probability goes to 1 for perturbation orders higher than 13. This actually results from the divergence of the perturbation expansion. It is legitimate, in these conditions, to analyze whether this divergence results from the Coulomb singularity at the nucleus or from  the infinite range of the potential. The Yukawa potential may be viewed as a screened Coulomb potential with a finite range which, in the present case, is equal to about 10 a.u. The results presented in Fig. \ref{fig5} show that the perturbation expansion diverges also in the 
case of the Yukawa potential. This seems to indicate that it is most likely the Coulomb singularity which is responsible for this divergence. Of course, this is not a rigorous mathematical proof 
of the existence and the origin of this divergence. It is important to study the behavior of this perturbation expansion for a broad range of frequencies and in particular in the low frequency 
regime. This latter problem which is much more demanding from the computational point of view will be addressed in a forthcoming publication.

\section{Results and discussions}
In this last section, we consider the interaction of atomic hydrogen with a 2-cycle pulse given by Eq. (\ref{eq_sin_pulse}). The frequency is equal to 0.057 a.u. that corresponds to a wavelength of 800 nm (Ti-Sapphire laser) and the peak intensity is 10$^{14}$ Watt/cm$^2$. As before, the carrier phase $\phi$ is set equal to zero. We calculate the electron energy distribution along the polarization axis both in the forward and backward
\begin{figure}[h]
\includegraphics[width=10cm,height=7.5cm]{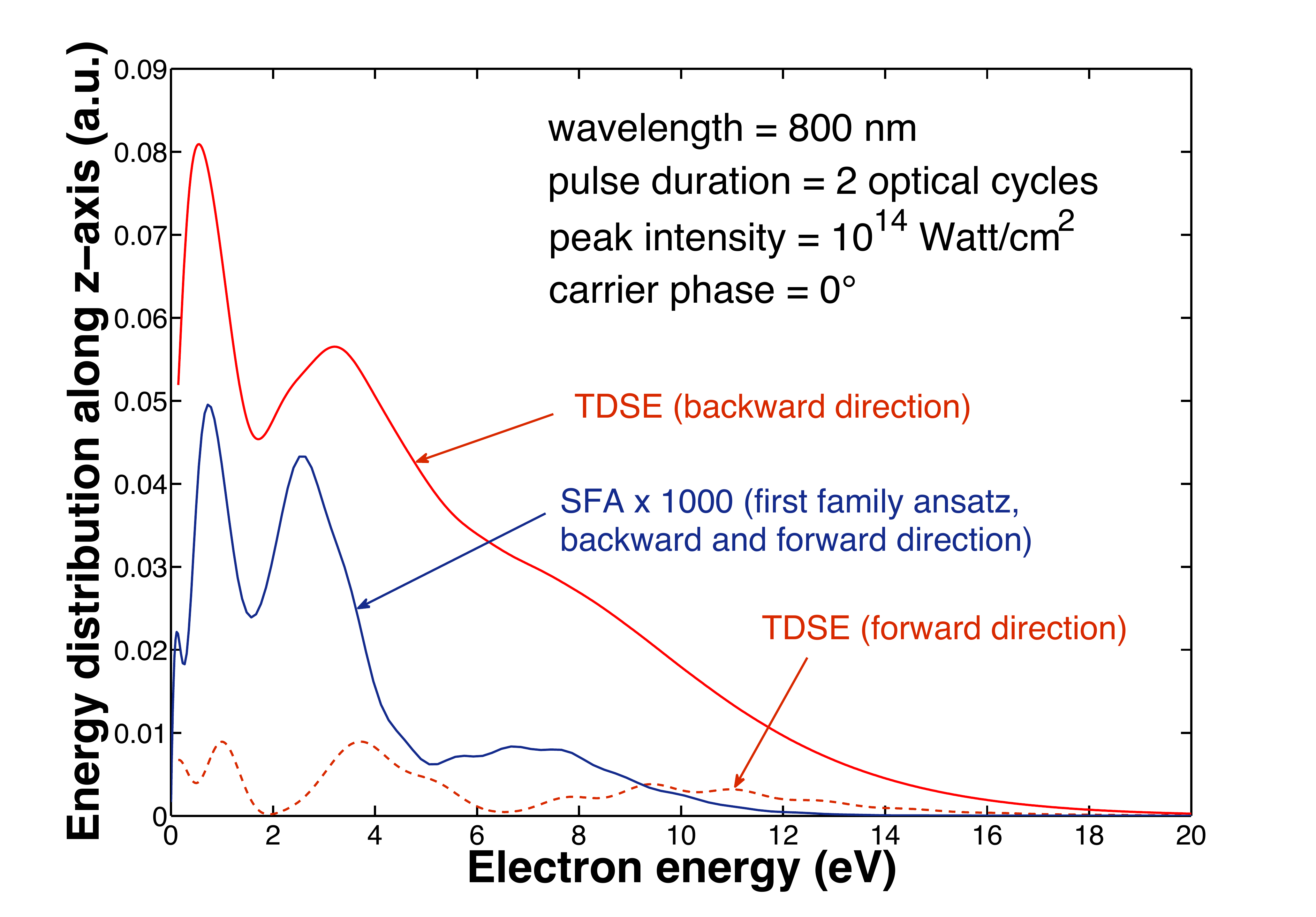}
\caption{(Color online) Energy distribution of the electrons emitted along the polarization axis, resulting from the interaction of atomic hydrogen with a 2-cycle pulse of frequency $
                \omega_0=0.057$ a.u. and peak intensity $I=10^{14}$ Watt/cm$^2$. The carrier phase is equal to 0 which means that the electric field exhibits a strong minimum at the middle of the
                pulse. The red curves correspond to the TDSE results in backward direction (full red line) and in the forward direction (red dashed line). The blue full line corresponds to the SFA 
                results (multiplied by a factor 1000) and obtained by using the ansatz (\ref{eq_phi1_V}) of the first family.}
\label{fig6}
\end{figure}
\begin{figure}[h]
\includegraphics[width=10cm,height=7.5cm]{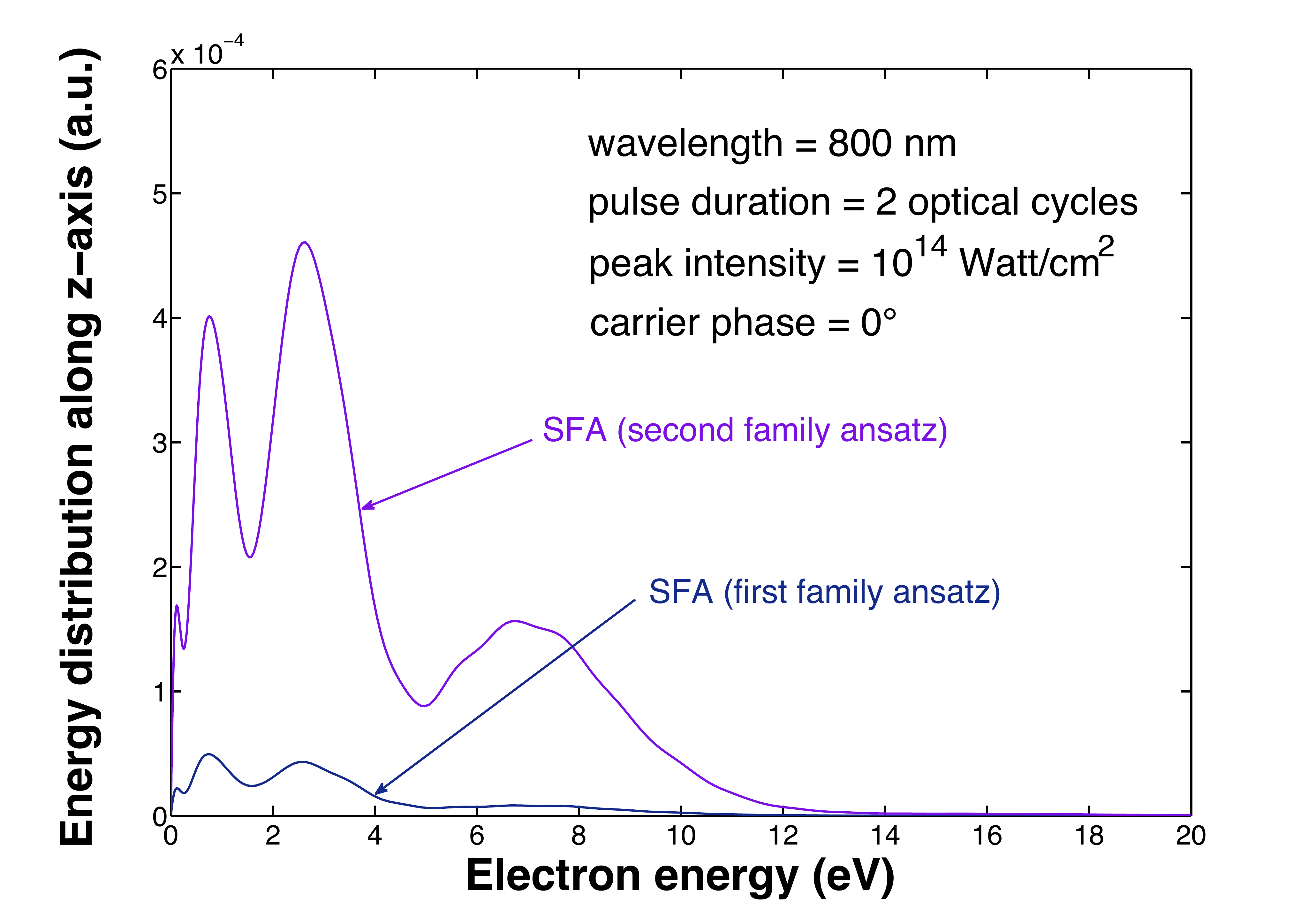}
\caption{(Color online) Energy distribution of the electrons emitted along the polarization axis, resulting from the interaction of atomic hydrogen with a 2-cycle pulse of frequency $
                \omega_0=0.057$ a.u. and peak intensity $I=10^{14}$ Watt/cm$^2$. The carrier phase is equal to 0. Both curves have been obtained within the SFA by using  ansatz (\ref{eq_phi1_V})
                of the first family (dark blue line) and ansatz (\ref{eq_phi2_L}) of the second family (magenta line).}
\label{fig7}
\end{figure}
\begin{figure}[h]
\includegraphics[width=12cm,height=7.5cm]{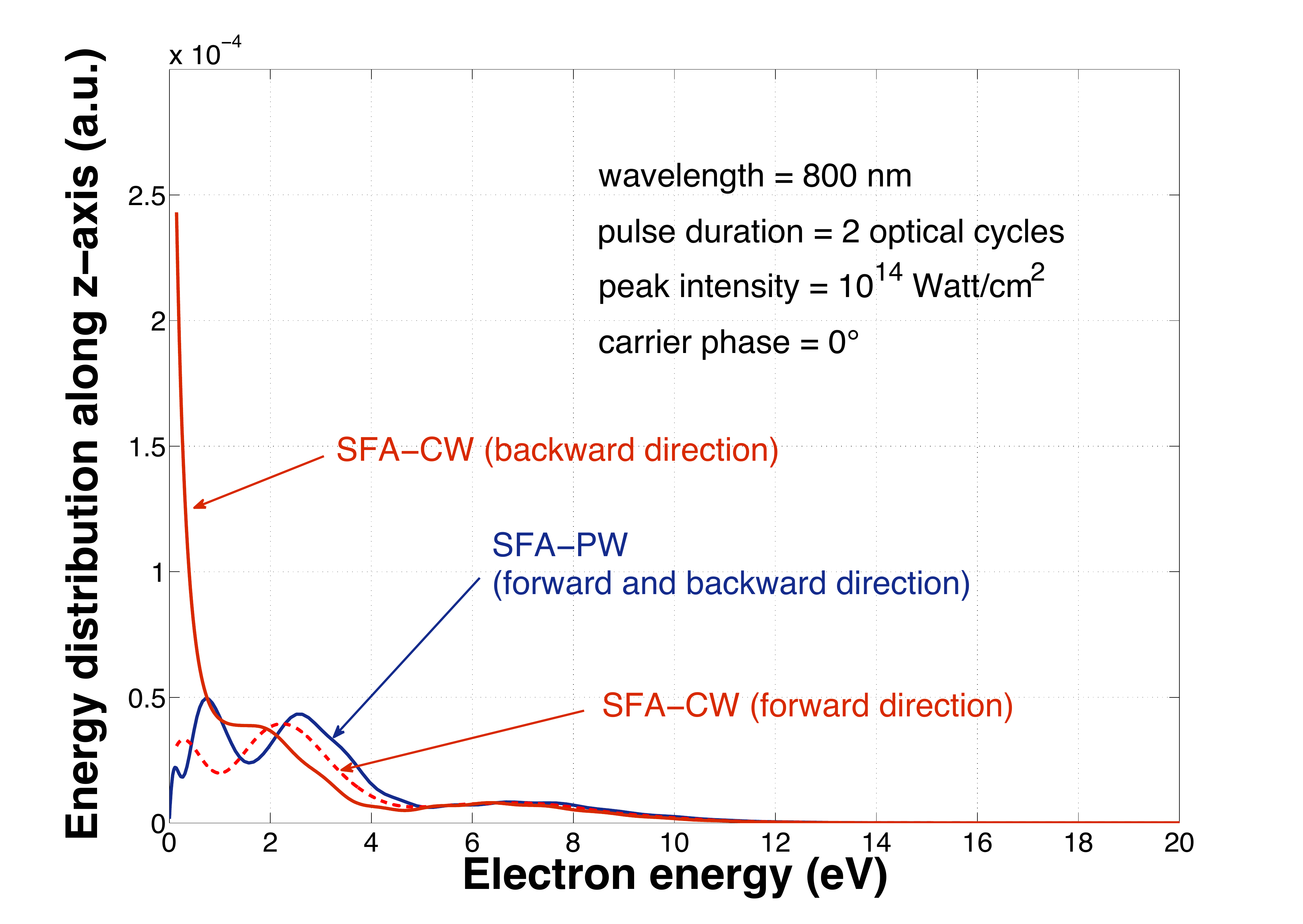}
\caption{(Color online) Energy distribution of the electrons emitted along the polarization axis, resulting from the interaction of atomic hydrogen with a 2-cycle pulse of frequency $
                \omega_0=0.057$ a.u. and peak intensity $I=10^{14}$ Watt/cm$^2$. The carrier phase is equal to 0. The full blue line corresponds to the SFA results obtained by using the V-gauge 
               ansatz of the first family and by projecting the final wave packet on plane waves. These SFA results are the same in the forward and backward directions. The red curves correspond to
               SFA results in which the final wave packet is projected on Coulomb waves. The red dashed curve refers to the forward direction and the full red curve to the backward direction.}
\label{fig8}
\end{figure}
\begin{figure}[h]
\includegraphics[scale=0.22]{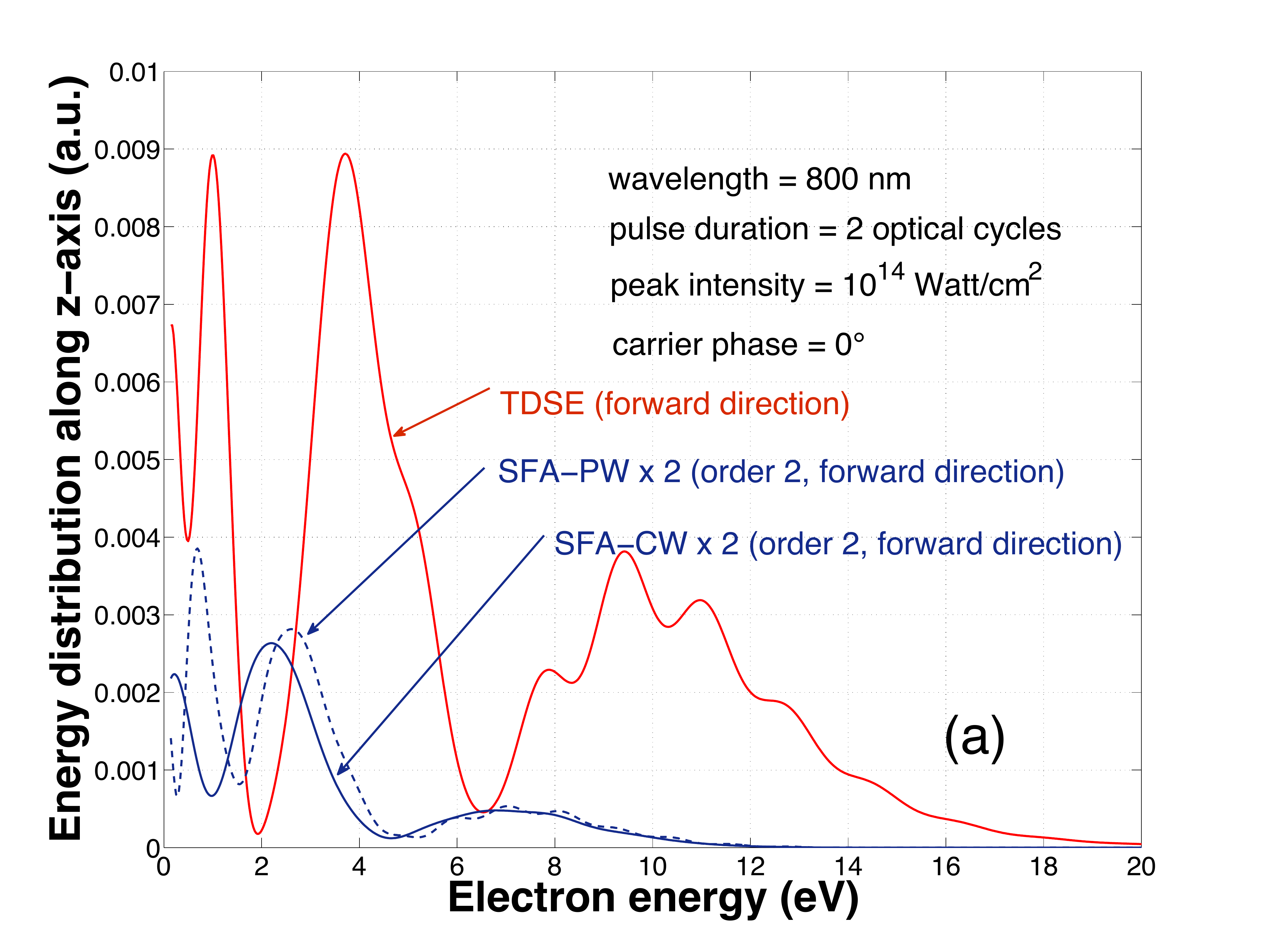}\includegraphics[scale=0.22]{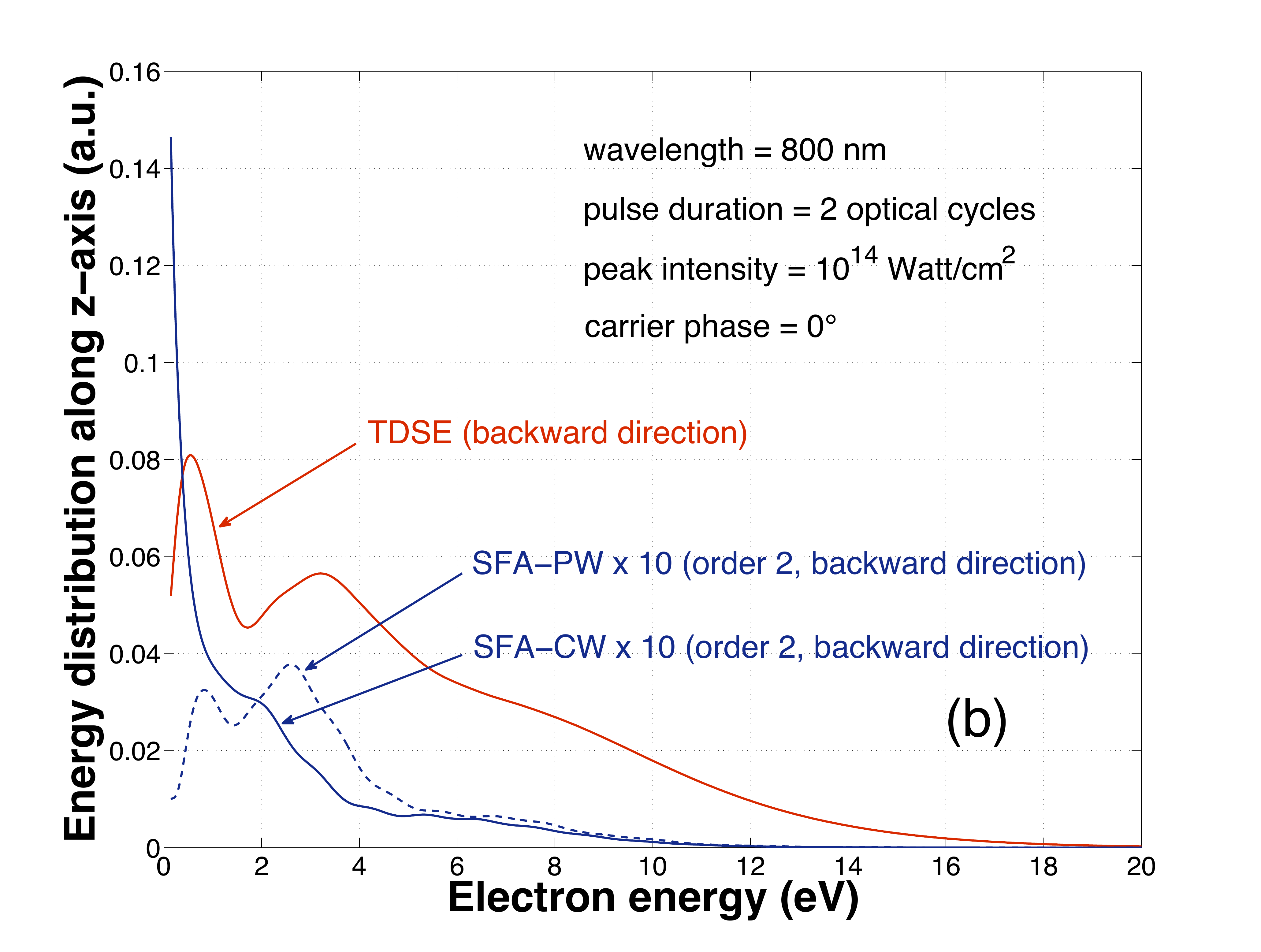}
\caption{(Color online) Energy distribution of the electrons emitted along the polarization axis in the forward (a) and in the backward (b) directions and resulting from the interaction of atomic
                hydrogen with a 2-cycle pulse of frequency $\omega_0=0.057$ a.u. and peak intensity $I=10^{14}$ Watt/cm$^2$. The carrier phase is equal to 0. The full red curves refer to the
                TDSE results. The blue curves correspond to the second order SFA results obtained by using the V-gauge ansatz of the first family. The blue dashed and full curves refer to these second 
                order SFA results obtained by projecting the final wave packet on plane waves and Coulomb waves respectively.}
\label{fig9}
\end{figure} 
direction and focus on the low energy part of this distribution. Our first objective is to compare TDSE to SFA results obtained by means of two ansatzes belonging to different families. The SFA calculations are performed by means of the semi-analytical formulae given in the previous section. Note that in these calculations, we explicitly calculate the norm of the SFA wave packet at the end of the interaction with the pulse. The second objective is to calculate the first and second order SFA wave packet fully numerically by using the V-gauge ansatz of the first family in the same physical situation. In this case also, the SFA results are fully normalized and the final SFA wave packets are projected on plane waves and Coulomb waves. \\

In Fig. \ref{fig6}, we show the TDSE results both in the backward and forward directions and the SFA results obtained by using ansatz (\ref{eq_phi1_V}). TDSE results show a striking dissymmetry between forward and backward directions. The main signal is observed in the backward direction because, as shown in \cite{hamidothesis}, the major part  of the emitted electronic wave packet undergoes a re-scattering by the ionic core in the backward direction. By contrast, the SFA calculations that  do not  take the re-scattering into account, give identical results in both directions. It is interesting to note that the SFA results are in good qualitative agreement with the TDSE results although three orders of magnitude lower. In Fig. \ref{fig7}, the V-gauge SFA result (first family) is compared to the L-gauge SFA result (second family). The overall behavior is qualitatively the same except that the L-gauge is one order of magnitude higher than the V-gauge result. The fact that  both L- and V-gauge (first order) SFA results are identical in the forward and backward directions may be explained from the semi-analytical formula provided that the SFA wave packet is projected on plane waves. In the following and before calculating the second order SFA amplitude, we examine whether this result is still valid if we project the (first order) SFA wave packet on Coulomb waves. In Fig. \ref{fig8}, we show the SFA results of a fully numerical calculation based on the V-gauge ansatz of the first family. When the SFA amplitude is projected on plane waves at the end of the pulse, we reproduce exactly the results of the calculation based on the semi-analytical formulae. In that case the energy distributions are the same in both directions. However, if we project the final SFA wave packet on Coulomb waves, we recover the strong dissymmetry observed in the TDSE results: electron emission in the backward direction is strongly favored with respect to the forward direction. However, the shape of the energy distribution in the backward direction is rather different from the corresponding TDSE result. It shows a sharp rise for electron energies going to 0. \\

In Fig. \ref{fig9}, we compare the TDSE results in both directions to second order SFA results that take into account  the re-scattering of the electrons by the ionic core. These second order SFA results are obtained fully numerically by using the V-gauge ansatz of the first family. They are properly normalized and the final wave packet is projected on plane waves and Coulomb waves. A first finding is that in the energy range considered here (below $2U_p$), including the second order contribution in the SFA amplitude is equivalent to multiplying the first order term by a constant factor, the value of which is very high. In the forward and backward directions, this factor is equal to about 500 and 100 respectively. In addition, this feature doesn't dependent on the
continuum wave function (plane wave or Coulomb wave) the final wave packet is projected on. This result confirms what we found in the previous section where we treated second order effects for 800 nm ($\omega_0=0.057$ a.u.) with an approximate calculation based on semi-analytical formulae (see Fig. \ref{fig1}) and at 65 nm ($\omega_0=0.7$ a.u.) with  a fully numerical calculation (see Fig. \ref{fig4}). In this context, it is important to note that the first order term and the second order term satisfy the inhomogeneous TDSE (60) and (61) respectively. Those two equations are the same except for the source terms. Beside this general remark, we also observe that in the forward direction, projecting the final wave packet on Coulomb waves instead of plane waves leads for energies below 5 eV, to a shift of the energy distribution towards lower energies. By contrast, in the backward direction, projecting on Coulomb waves drastically changes the shape of the energy distribution.\\

\section{Conclusion and perspective}
In this contribution, we showed that the various treatments based on the strong field approximation may be grouped into a set of families of approximation schemes. We introduced different ansatzes that describe the electron wave packet as the sum of the initial state wave function multiplied by a phase factor and a function which is the perturbative solution in the Coulomb potential of an inhomogeneous TDSE. It is the phase factor that determines which family of approximation schemes the SFA treatment belongs to. We considered two families. In each of them, the velocity and the length gauge versions of the approximation scheme give, by construction, identical results at each order in the Coulomb potential. By contrast, and  irrespective of the gauge, approximation schemes belonging to different families  lead to different results. We showed how to construct and solve numerically the SFA wave packet to arbitrary order using an iterated inhomogeneous TDSE. This also allowed us to project the solution onto either Coulomb or plane wave final states.\\

This reformulation clarifies the long standing problem of the gauge in the context of the SFA. It also allowed us to address two  important issues. The first one concerns the role of the Coulomb potential in the output channel. In other words, if we project the SFA wave packet at the end of the interaction with the external field on Coulomb waves instead of plane waves as it should be, how this will affect the low energy part of the electron energy distribution? The second issue is the convergence of the perturbative series in the Coulomb potential. In order to study the role of the Coulomb potential in the output channel, we used the V-gauge ansatz of the first family. In that case, the inhomogeneous TDSE can be solved numerically by means of a spectral method. At high frequency, all the low energy structures are shifted towards lower energies as expected. At low frequency however, drastic changes of the shape of the energy distribution occur. To address the problem of the convergence of the perturbative series, we first studied the second order SFA for a wavelength of 800 nm with two different methods. The first one is based on approximate semi-analytical formulae associated to the L-gauge ansatz of the second family. The second method is based on the fully numerical solution of the inhomogeneous TDSE associated to the V-gauge ansatz of the first family. Both methods show that for electron energies  below $2U_p$, the second order SFA amplitude is actually the first order one multiplied by a large factor. By using the second method in a much simpler case namely a frequency of 0.7 a.u. we were able to calculate many high order terms. Our results indicate that in this particular situation, the perturbative series actually 
diverges. Of course, this does not represent a rigorous proof of the divergence of this series. This point which deserves a deeper analysis will be treated elsewhere.

\section{Acknowledgements}
A.G. is "aspirant au Fonds de la Recherche Scientifique (F.R.S-FNRS)". The authors are very grateful to Misha Ivanov and Olga Smirnova for very enlightening discussions.
Y.P. and F.C. thanks the Universit\'e Catholique de Louvain (UCL) for financially supporting several stays at the Institute of Condensed Matter and Nanosciences of the UCL. F.M.F and P.F.O'M 
gratefully acknowledge the European network COST (Cooperation in Science and Technology) through the Action CM1204 "XUV/X-ray light and fast ions for ultrafast chemistry" (XLIC) for financing several short term scientific missions at UCL. The present research benefited from computational resources made available on the Tier-1 supercomputer of the F\'ed\'eration Wallonie-Bruxelles funded by the R\'egion Wallonne under the grant n$^o$1117545 as well as on the supercomputer Lomonosov from Moscow State University and on the supercomputing facilities of the UCL and the Consortium des Equipements de Calcul Intensif (CECI) en F\'ed\'eration Wallonie-Bruxelles funded by the F.R.S.-FNRS under the convention 2.5020.11. A.G. and Y.P. are grateful to Russian Foundation for Basic Research (RFBR) for the financial support under the grant N14-01-00420-a. F.C. and B.P. also thank l' Agence Nationale de la Recherche française (ANR) in the context of «Investissements d'avenir» Programme IdEx Bordeaux - LAPHIA (ANR-10-IDEX-03-02) and Computer time for this study that was provided by the computing facilities MCIA (M\'esocentre de Calcul Intensif Aquitain) of the Universit\'e de Bordeaux and of the Universit\'e de Pau et des Pays de l'Adour.

\newpage

\end{document}